\documentclass[epjCONF]{svjour}
\usepackage[varg]{txfonts} % Times fonts
\usepackage[latin1]{inputenc}
\usepackage{xspace}
\usepackage[sc]{mathpazo}
\usepackage{graphicx}
\usepackage{units}
\usepackage{url}
\usepackage{color}
\usepackage{subfigure}

\session-title{UHECR2012: International Symposium on Future Directions in UHECR 
Physics}
\def\Xmax{$X_\mathrm{max}$\xspace}%
\def\sigmaXmax {RMS$(X_\mathrm{max})$\xspace}%
\def\sigmaHiRes {$\sigma_{X}$\xspace}%
\def\meanXmax {$\langle X_\mathrm{max}\rangle$\xspace}
\def\meanXmaxBias{$\langle X_\mathrm{max}^\mathrm{meas}\rangle$\xspace}%
\def\eV {\textrm{e\kern -0.1em V}\xspace}%
\def\gcm{g/cm$^2$\xspace}%
\def\meanlnA{$\langle \mathrm{lnA} \rangle$\xspace}%
\def\FOVmin{X$^\mathrm{min}_\mathrm{fov}$\xspace}%
\def\FOVmax{X$^\mathrm{max}_\mathrm{fov}$\xspace}%
\newcommand{\energy}[1]{\unit[$10^{#1}$]{\eV}}

\begin{document}

\title{Mass Composition Working Group Report}
\author{E.~Barcikowski\inst{1} \and J.~Bellido\inst{2}\thanks{jbellido@physics.adelaide.edu.au} \and J.~Belz\inst{1} \and Y.~Egorov\inst{3} \and S.~Knurenko\inst{3} \and V.~de Souza\inst{4} \and Y.~Tameda\inst{5} \and Y.~Tsunesada\inst{6} \and M.~Unger\inst{7}
for the HiRes, Pierre Auger, Telescope Array and Yakutsk Collaborations}

\institute{Physics Department, University of Utah, USA \and 
University of Adelaide, Adelaide, S.A. 5005, Australia \and 
Yu. G. Shafer  Institute for Cosmophysical Research and Aeronomy, 31 Lenin Ave.,
 677980 Yakutsk, Russia \and 
Universidade de S\~ao Paulo, Instituto de F\'{\i}sica de S\~ao Carlos, SP, Brazil \and 
Institute for Cosmic Ray Research, the University of Tokyo,
Kashiwa, Chiba 277-8582, Japan \and 
Graduate School of Science and Engineering, Tokyo Institute of Technology, 
Meguro, Tokyo, Japan 152-8550 \and 
Karlsruhe Institute of Technology, Campus North, Institut f\"ur Kernphysik, Karlsruhe, Germany}
\abstract{
  We present a summary of the measurements of mass sensitive parameters
  at the highest cosmic ray energies done by several experiments.
  The \Xmax distribution as a function of energy has been
  measured with fluorescence telescopes by the HiRes, TA  and Auger
  experiments and with Cherenkov light detectors by Yakutsk.
  The \meanXmax or the average mass (\meanlnA) has been also inferred using
  ground detectors, such as muon and water Cherenkov detectors.
  We discuss the different data analyses elaborated by each
  collaboration in order to extract the relevant information. Special
  attention is given to the different approaches used in the
  analysis of the data measured by fluorescence detectors in order to
  take into account  detector biases. We present a careful analysis
  of the stability and performance of
  each analysis. The results of the different experiments will be
  compared and the discrepancies or agreements will be  quantified.
} %end of abstract
%
%\linenumbers\relax
\maketitle
\section{Introduction}
\label{intro}

In preparation for this meeting, several working
groups were formed to establish a common view on the experimental
status of measurements at ultra-high energies. Here we report the
findings of the {\itshape mass composition working group} consisting
of members of the Auger, HiRes, Telescope Array (TA) and Yakutsk 
collaborations. The
aim was to understand and quantify potential differences between
different experimental results and to try to discuss the measurements
in terms of the cosmic ray mass composition.

A current issue in the field of ultra high energy cosmic rays is that
the measurements~\cite{Abraham} of the depth of shower maximum from
Auger are shallower and less fluctuating than predictions from air
shower simulations for a pure proton composition at energies $\gtrsim$
\energy{19}. On the other hand, the HiRes and TA
results~\cite{bib:hires,bib:ta} in the same energy regions are
 consistent with QGSJet-II simulations assuming a constant
composition dominated by light elements.

However, there are important differences in the analysis of the data
between Auger, HiRes and TA. In this note we will review the
differences between the analyses, compare the different experiment
results, and evaluate their compatibility within the
experimental statistical and systematic uncertainties.

The main focus of this work is the understanding of the apparent
differences between the Auger, HiRes and TA \Xmax
results. Nevertheless, we also discuss in this paper other shower
observables which are sensitive to the mass of the primary cosmic
rays. The Yakutsk experiment has muon detectors operating and this
information is used to infer \meanlnA in a complementary
way~\cite{bib:yakutsk:muons}. In a similar way, the Auger
collaboration operates an array of water Cherenkov
detectors~\cite{bib:auger} that are able to measure the muonic and
electromagnetic signal at the ground level. This signal is used to
infer \meanlnA.

The challenge for determining the cosmic ray mass composition at the
highest energy is our limited knowledge of the hadronic interaction
properties. The current hadronic interaction models extrapolate
interaction properties measured in particle accelerators at energies
more than two orders of magnitude smaller. If we knew the hadronic
interaction properties at these higher energies, the interpretation of
the data in terms of the primary cosmic ray composition would be
straightforward. Despite this limitation, we used the hadronic 
interaction models to convert the different observables measured by each 
experiment into \meanlnA to allow direct comparison of the measurements.

%In the following sections, we review the \Xmax
%analysis (Sec.~\ref{sec:analysis}), show the measurements (Sec.~\ref{sec:xmax})
%and its stability (Sec.~\ref{sec:stability}), and compare the different
%experiments (Sec.~\ref{sec:xmaxComparisons}). In Section~\ref{sec:other:par}
%we discuss other mass composition parameters. Section~\ref{sec:discussion}
%contains the conclusions of the paper.

\section{Different Approaches to Analyse \Xmax Observations}
\label{sec:analysis}

The fluorescence detectors (FDs) of the Auger~\cite{bib:auger:fluor},
HiRes~\cite{bib:HiRes:fluor} and TA~\cite{bib:ta:fluor} experiments have
a limited field of view (FOV) ranging from about 3$^\circ$ to 30$^\circ$ in elevation.
This limited FOV introduces a detector bias depending on the shower
geometry: For close by showers, the lack of a high elevation FOV
prevents the observation of  shallow showers and, therefore, deeper showers
are favored in the event selection. Moreover, deep near-vertical
showers may reach their maximum below the FOV range (or even
below ground level) and therefore the ensemble of selected showers is
artificially enriched with shallow showers.  To address this problem
the collaborations follow two different approaches.

\subsection{The Auger Approach}
\label{auger:approach}

The goal of the Auger Collaboration is to publish \meanXmax and
\sigmaXmax values with minimal detector bias, that are close to the
moments of the undistorted distribution.  This procedure is only
limited by systematic uncertainties.  The advantage of this proposal
is that the published results can be compared directly with model
expectations, without the need of any knowledge of the detector and,
therefore, also to simulations with hadronic interaction models and or
composition hypotheses that were not available at the time of
publication.

In order to measure an unbiased \Xmax distribution, it is necessary to
select events based only on their arriving geometry and energy
(i.e.\ not using any information of their \Xmax value). The basic idea
is to select showers with geometries that will allow \Xmax to be inside a
detector FOV that is ``wide enough'' to cover the true \Xmax
distribution.
In order to choose the appropriate geometries to
be considered in the \Xmax analysis, it is necessary to have a rough
idea of what the range of the \Xmax distribution is. The idea is to use
the data themselves to determine this range for each
energy bin.

Given the geometry and energy of an event, the range in \Xmax for which a
shower can be detected with good quality can be predicted with a semi-analytical calculation that takes into account the atmospheric attenuation and the optical efficiency of the detector. The result
of such a calculation is an \Xmax-range, \FOVmin to \FOVmax, for which a shower will be accepted for the data analysis defining the {\itshape effective} (as opposed to geometrical) FOV of the telescopes.

\begin{figure}[h]
\includegraphics[width=0.95\linewidth]{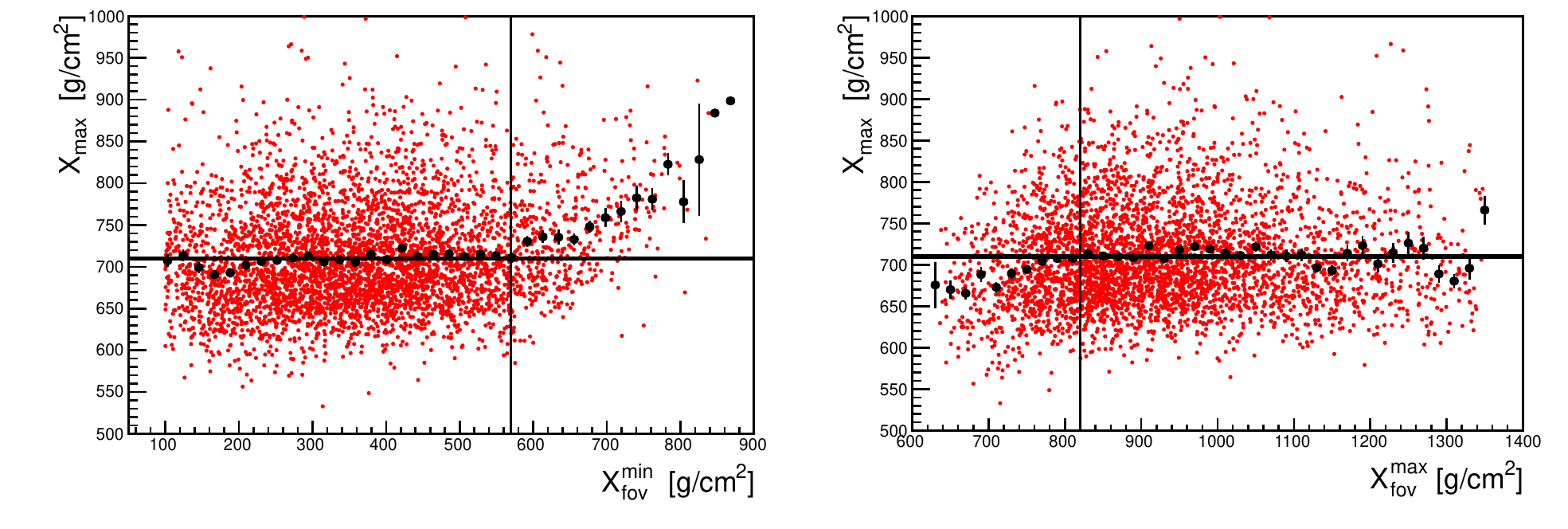}
\caption{Scatter plots of the measured \Xmax between $10^{18.0}$ and \energy{18.2} as a function of the
  event-by-event field of view boundaries \FOVmin and \FOVmax.  The
  mean \Xmax values as a function of \FOVmin and \FOVmax are
  superimposed as large black dots. The horizontal lines denote the
  asymptotic \meanXmax value far away from the field of view boundary.
  The range where the \meanXmax values are consistent with this
  line defines the unbiased region. The vertical lines indicate the
  limits of fiducial field of view that result in an unbiased \Xmax
  measurement.}
\label{fig:FOV}
\end{figure}

Fig.~\ref{fig:FOV} shows the scatter plots of the measured \Xmax
between $10^{18.0}$ and \energy{18.2} as a function of the estimated
\FOVmin and \FOVmax as well as the dependence of the average \Xmax
values on these variables. The two plots clearly illustrate the
resulting bias that is introduced when \FOVmin is too deep (panel on
the left) or when \FOVmax is too shallow (panel on the right).  The
Auger Collaboration analysis procedure uses these graphs for each
energy bin in order to determine what ranges of \FOVmin and \FOVmax
values allow for an unbiased sampling of the \Xmax distributions as
a function of energy~\cite{bib:unger:xmax,bib:bellido:xmax}.  The vertical
lines in Figure~\ref{fig:FOV} indicates the appropriate limits for
\FOVmin and \FOVmax. These limits determine the fiducial field-of-view cuts. 
These fiducial field-of-view cuts are optimized independently for the data 
and for each Monte Carlo (MC) composition. Different MC compositions 
(i.e. different \Xmax distributions) have been used to test this 
algorithm~\cite{bib:unger:xmax,bib:bellido:xmax} and the 
reconstructed \meanXmax values were found consistent with the MC input within 
statistical uncertainties.

The Auger collaboration applies identical quality cuts to data and 
Monte Carlo, which is similar to the approach shared by HiRes and TA 
(in the analysis described below) in which equivalent cuts are applied to 
data and Monte Carlo throughout. The additional field-of-view cuts --- for 
which there is no direct analog in the HiRes or TA analysis ---  are optimized 
independently for the data and for each MC composition, using the same 
algorithm for this optimization. The motivation behind 
this choice is to allow optimization of the field-of-view cuts without making 
any {\em a~priori} assumptions as to the range of the $X_{max}$ distribution.

Above \energy{18.2}, the application of the  field-of-view cuts reduce the Auger statistics by half (Table~\ref{tab:statistics}). The severity of these  field-of-view cuts (for reducing the statistics) puts a constraint on the minimum number of events required for this approach.  

\subsection{The HiRes and TA Approach}
\label{hires:approach}

 The HiRes and TA collaborations do not apply field-of-view cuts. This 
means that they do not attempt to measure the unbiased \meanXmax  
in the  atmosphere, instead they quote the \meanXmax as 
measured in the detector (we will refer it as \meanXmaxBias). These two 
\meanXmax values can differ significantly depending on the intrinsic \Xmax 
distribution, or depending on the zenith angle distribution of the events 
(as shown in Figs.~\ref{fig:hires:xmax:vert}~and~\ref{fig:hires:xmax:incl}). 
However, this should not affect the HiRes/TA \Xmax composition analysis, 
because an accurate detector modeling is used for predicting the \meanXmaxBias 
observations for a given composition.

In order to understand the acceptance and reconstruction biases arising from the inherent field-of-view limitations of a fluorescence detector, both the HiRes and TA collaborations focus on accurate detector modeling through the use of a detailed detector Monte Carlo. Air showers are generated using CORSIKA~\cite{bib:corsika}  and several hadronic interaction models including QGSJet01~\cite{bib:qgs01}, QGSJet-II~\cite{bib:qgsII} and SIBYLL~\cite{bib:sib,bib:sib2}. Shower libraries are created in which the number of particles as a function of slant depth is recorded for a large number of air showers induced by different primary masses. 

In the detector simulation, an event is drawn from the library and assigned a random core location, zenith and azimuthal angle. The fluorescence light is propagated from the shower to the detector, with attenuation simulated via the use of an empirically determined atmospheric database. Ray tracing is performed to determine the photoelectron response of individual PMT's in the fluorescence camera. The trigger algorithms are simulated, and if the trigger conditions are satisfied the Monte Carlo event is written to disk in the identical format as real data, allowing study by the same analysis chain. 

A number of control distributions are checked for agreement between data and Monte Carlo in order to assure that all detector effects are accurately described by the simulation. Particular attention is paid to those distributions --- {\em e.g.} first and last viewed depth of the shower --- which touch closely on the detector biasing issue. Finally, to extract information about composition the observed $X_{max}$ distributions in the data are compared to the "observed" $X_{max}$ distributions for Monte Carlo events which pass identical event selection criteria.

In summary, in order to estimate an average mass composition, the
\meanXmax measured by Auger can be compared directly with the
predictions from air shower simulations (within remaining systematic uncertainties). In the case of HiRes and TA,
the measured \meanXmaxBias should be compared with the \meanXmaxBias
obtained from a convolution of simulated showers with a model of the
detector, atmosphere and reconstruction.

It is important to note that the \meanXmax measurements by Auger and Yakutsk
cannot be compared directly with the \meanXmaxBias published by HiRes
and TA. In Sec.~\ref{sec:xmaxComparisons} we will transform these \meanXmax and \meanXmaxBias to \meanlnA to compare the different experiment results.

\section{Yakutsk Measurement of the \Xmax distributions}

The determination of \Xmax in individual showers is based on the measurement
of the Cherenkov light flux at different core distances Q(r):
\begin{enumerate}
 \item by the parameter p = lg(Q(200)/Q(500));
 \item reconstruction of a shower development curve from the lateral 
distribution of Cherenkov light~\cite{bib:Knurenko};
  \item measurement of the Cherenkov light pulse width at a fixed core distance;
  \item by recording a Cherenkov track with a differential detector based on
     camera obscura.
\end{enumerate}

The sensitivity of these techniques is described in~\cite{bib:Hillas}.
The accuracy of \Xmax determination in individual showers was estimated in
a simulation of EAS characteristics measurements at the array involving
MC methods and amounted to 30-45~\gcm , 35-55~\gcm, 15-25~\gcm,
35-55~\gcm respectively for the first, second, third and fourth methods.
The total error of \Xmax estimation included errors associated with core
location, atmospheric transparency during the observational period, hardware
fluctuations and mathematical methods used to calculate a main parameters.

\section{Comparing Different \Xmax Measurements}
\label{sec:xmax}

\begin{figure}[h]
\center{\includegraphics[width=0.75\linewidth]{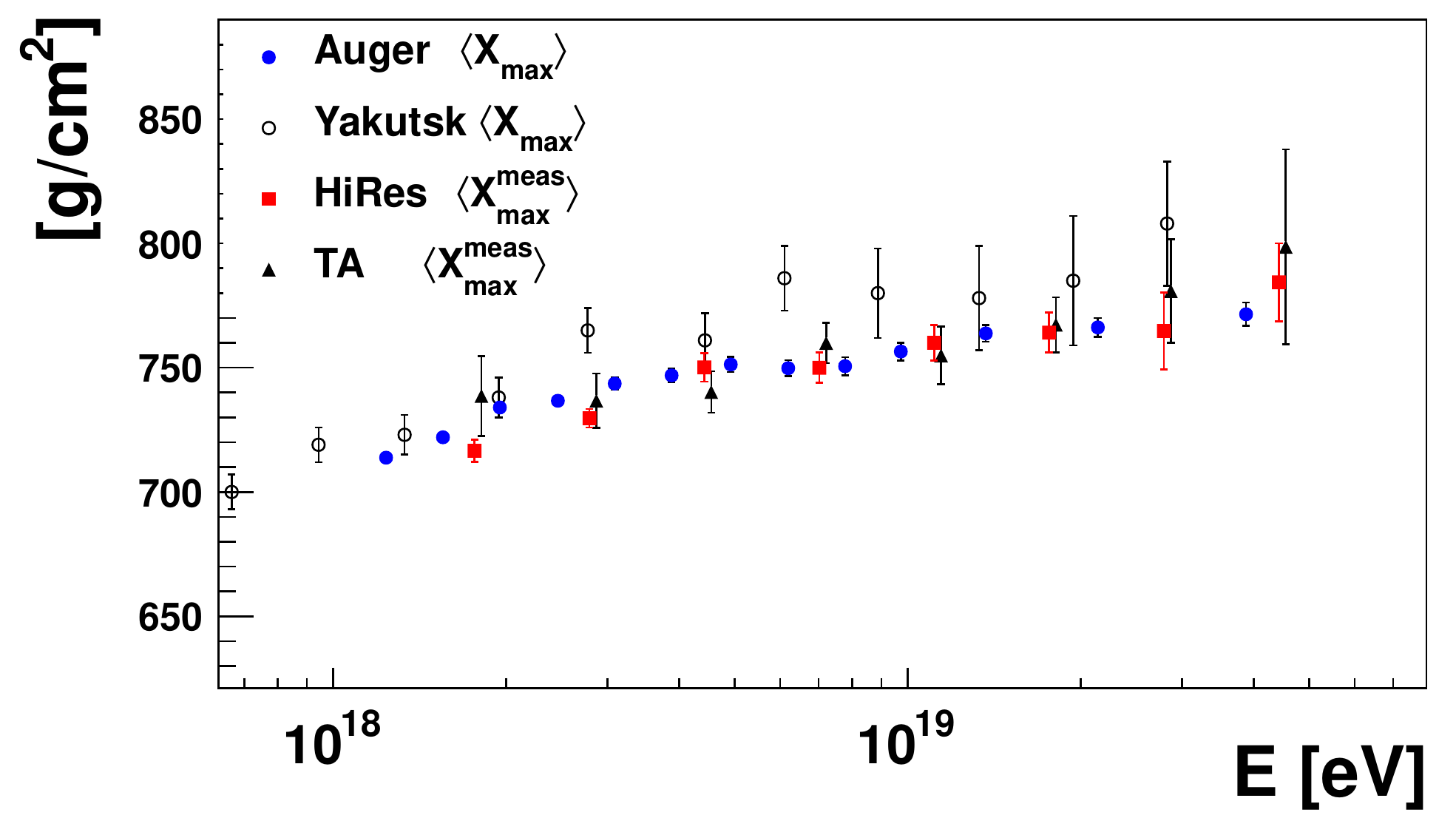}}
\caption{ \meanXmax measured by Auger and Yakutsk, together with the
  \meanXmaxBias as measured by HiRes and TA.  Data points are shifted
  to a common energy scale (text for details).}
\label{fig:all:xmax}
\end{figure}

Fig.~\ref{fig:all:xmax} shows the \meanXmax measured by
Auger~\cite{bib:auger:xmax:icrc11} and
Yakutsk~\cite{bib:yakutsk:xmax}, together with the \meanXmaxBias as
measured by HiRes~\cite{bib:hires} and
TA~\cite{bib:ta}. The observed agreement between the measured
\meanXmax and \meanXmaxBias is not expected.

At this meeting, the energy spectrum working group has compared the
shape of the energy spectrum from the Auger, Yakutsk, HiRes and TA
experiments and has produced a table with normalization
factors~\cite{bib:WG:spectrum}. For the plots presented here, we have
normalized the energy scales to an energy scale that is half way
between the Auger and TA energy scales. The normalization factors that
we have used are 1.102 for Auger, 0.55 for Yakutsk, 0.883 for HiRes
and 0.908 for TA. Later in Sec.~\ref{sec:xmaxComparisons} we will
evaluate the compatibility of the different results. We will transform
\meanXmax and \meanXmaxBias to \meanlnA for meaningful comparisons.

\begin{figure}[h]
\center{\includegraphics[width=0.95\linewidth]{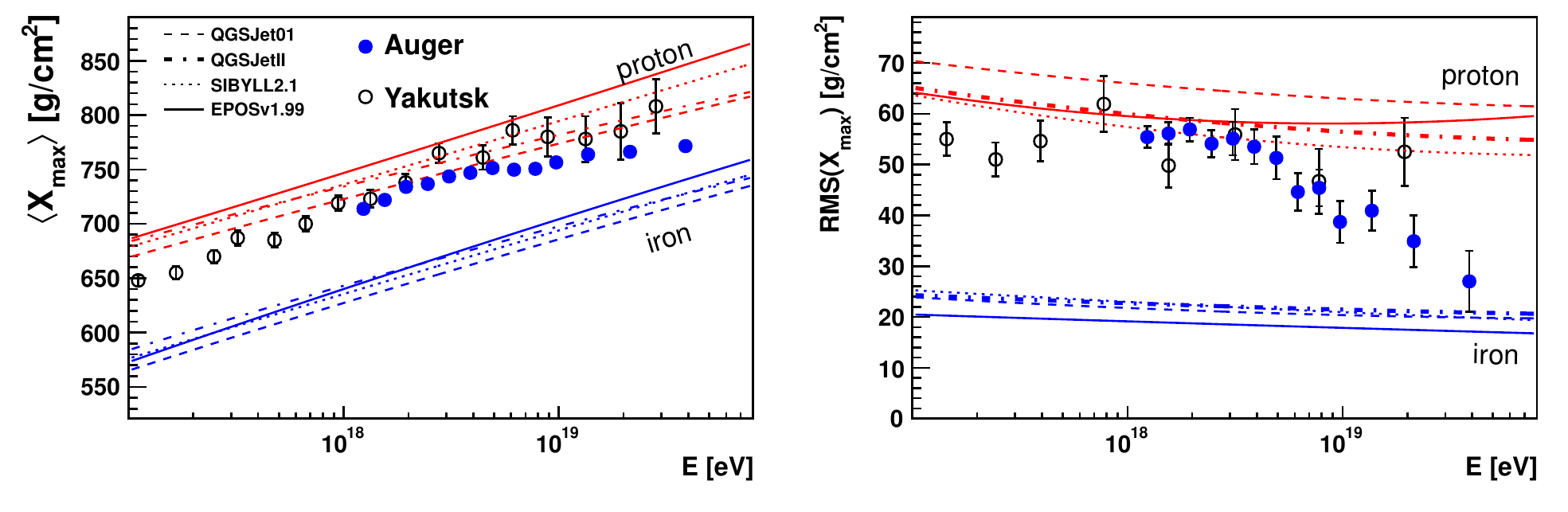}}
\caption{Measured \meanXmax (left) and \sigmaXmax (right) for the Auger
  and Yakutsk experiments. The lines indicate the \meanXmax
  expectations for proton and iron compositions using different
  hadronic interaction models. Notice that the highest energy bin for Yakutsk contains \textbf{only 3} events (Fig.~\ref{fig:xmax:statistics}).}
\label{fig:auger:yakutsk:xmax}
\end{figure}

Fig.~\ref{fig:auger:yakutsk:xmax} shows the measured \meanXmax (panel
on the left) and \sigmaXmax (panel on the right) for the Auger and Yakutsk
experiments. Since both experiments published \meanXmax values with
minimum detector bias, we can compare both of them with the model
expectations. The same holds true for the measurements of the
shower-to-shower fluctuations, where both experiments corrected the
measurements for the detector resolution.  The lines indicate the
predictions from air shower simulations for proton and iron
compositions. There are different line types corresponding to
different high energy hadronic interaction models: QGSJet-01, QGSJet-II,
SIBYLL2.1 and EPOSv1.99.

\begin{figure}[h]
\includegraphics[width=0.48\linewidth]{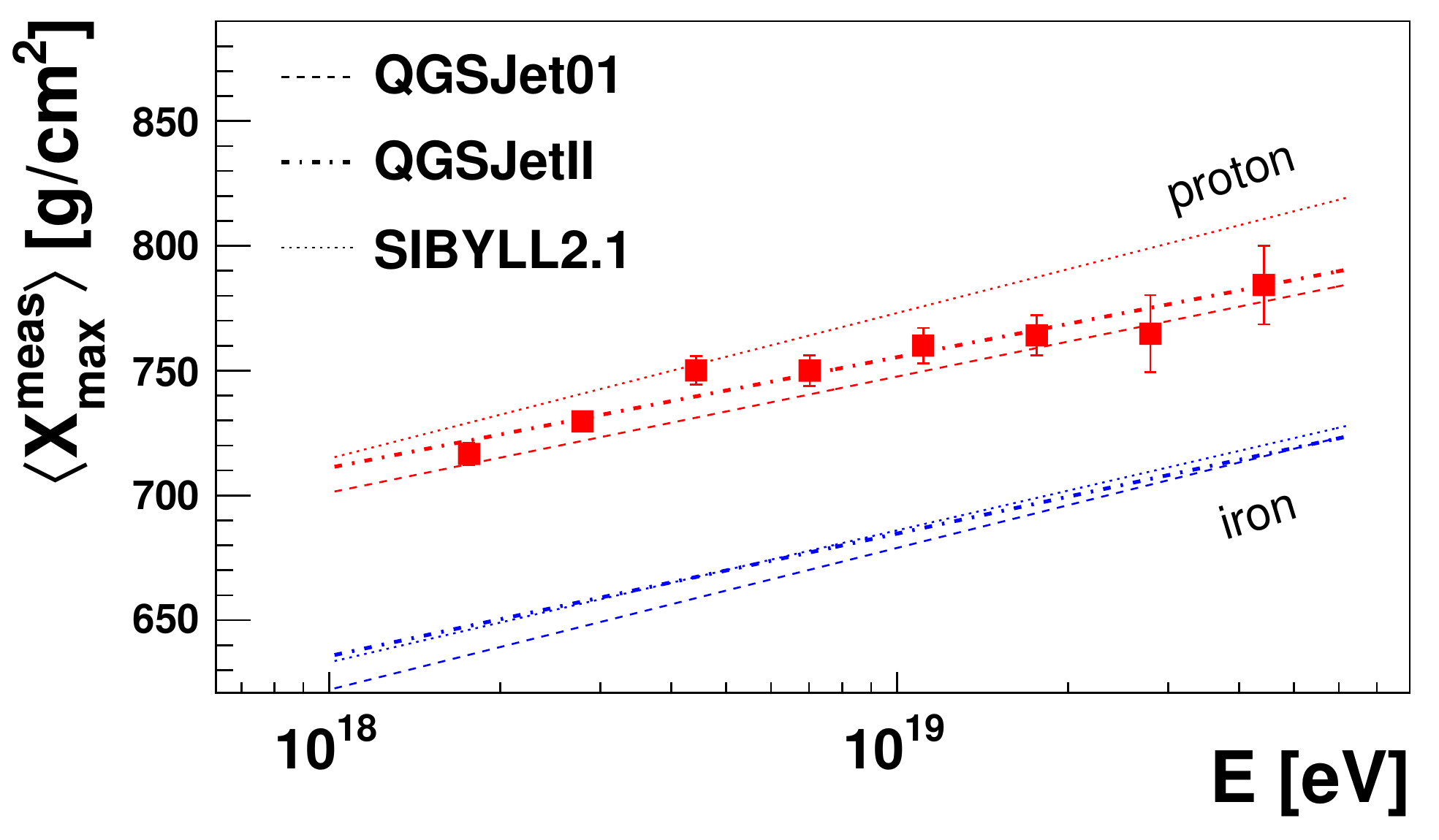}
\includegraphics[width=0.48\linewidth]{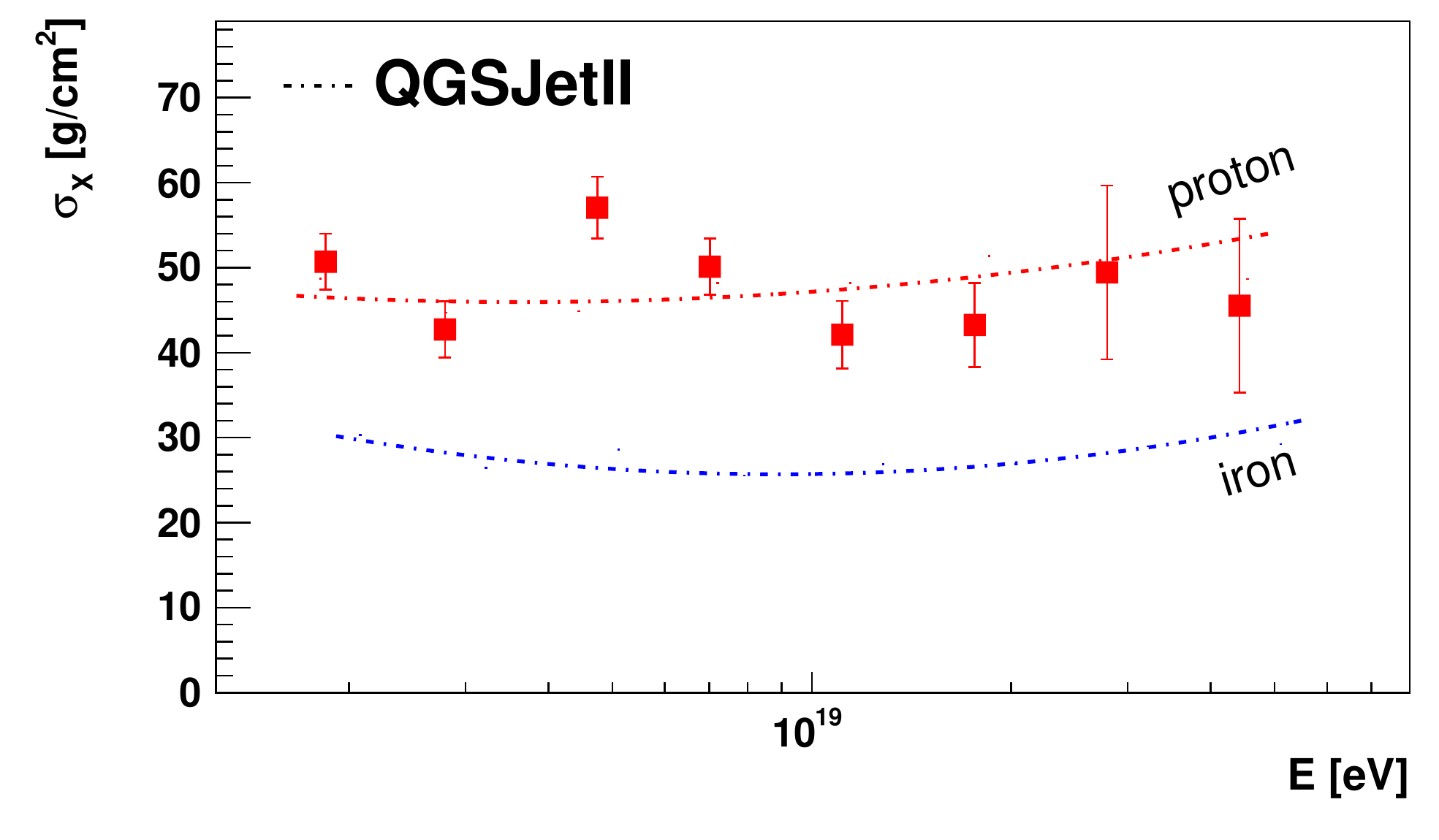}
\caption{The \meanXmaxBias (left) and \sigmaXmax (right) as measured
  by the HiRes experiment. The lines are the corresponding
  \meanXmaxBias and \sigmaHiRes expectations for proton and iron
  compositions. The different line types correspond to different
  models.}
\label{fig:hires:xmax}
\end{figure}

%----------------------------------------------------
\begin{figure}[h]
\begin{minipage}[b]{0.5\linewidth}
  \begin{center}
      \includegraphics[width=0.9\linewidth]{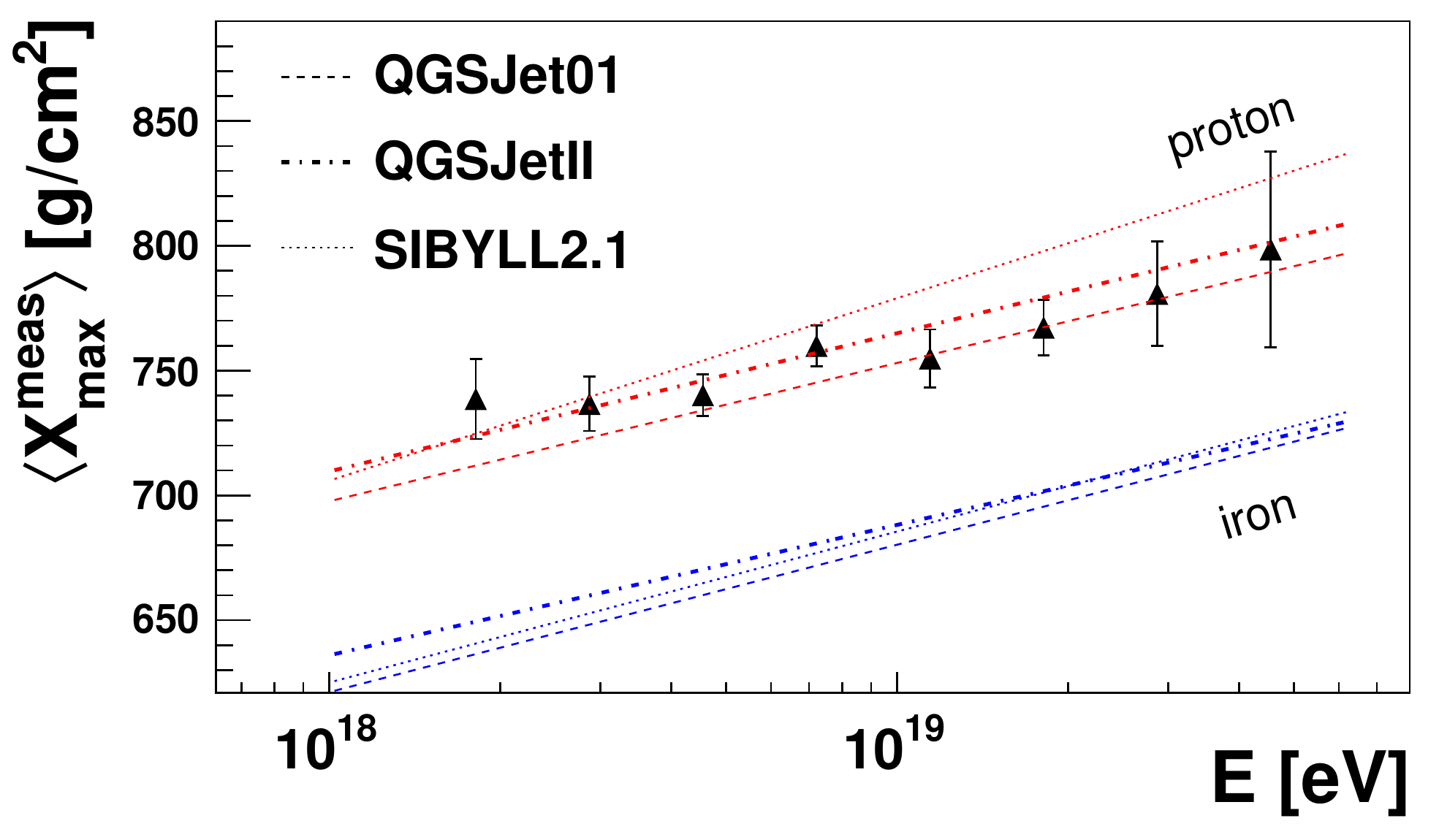}
      \caption{The \meanXmaxBias measured by the TA experiment. The
        lines are the corresponding \meanXmaxBias expectations for
        proton and iron primaries. The different line types correspond
        to different models.}
      \label{fig:ta:xmax}
  \end{center}
\end{minipage}
\hspace{0.5cm}
\begin{minipage}[b]{0.45\linewidth}
  \begin{center}
    \includegraphics[width=0.9\linewidth]{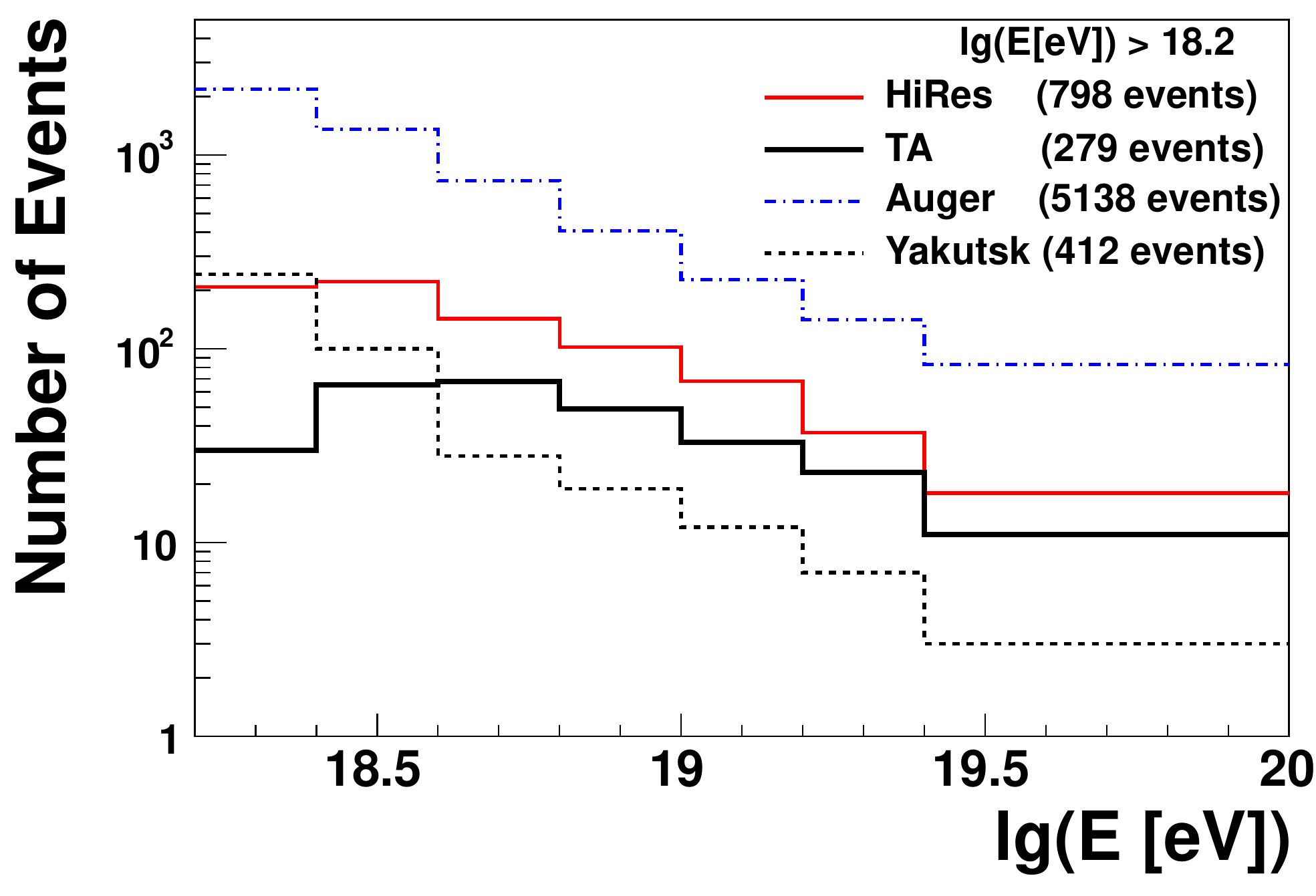}
    \caption{Number of events that survived the selection cuts as a
      function of energy. For this plot the energies have been normalized to the TA energy scale.}

      \label{fig:xmax:statistics}
  \end{center}
\end{minipage}
\end{figure}

\begin{table}[h]
  \centering
  \begin{tabular}{|c|c|c|c|c|c|c|} \hline
             &  Auger         & Auger            & HiRes         & HiRes & TA  & Yakutsk \\
             &  standard cuts & without FOV cuts & standard cuts & no Rp cuts  &  &  \\ \hline
E $>$ \energy{18.2} & 5138   &  11343  & 798 & 1306 & 279 & 412 \\ \hline
E $>$ \energy{19.0} & 452    &  709    & 123 & 143  & 67  &  22 \\ \hline
%Probability  & $8.48\times 10^{-24}$ & 0.72 & 0.19 & 0.35 \\ \hline
  \end{tabular}
  \caption{Number of events above  \energy{18.2} and \energy{19.0} (Fig.~\ref{fig:xmax:statistics}). In this table we have included the number of reconstructed Auger events that survived all the quality cuts (i.e. number of events prior to the application of the field-of-view cuts). The energy distribution of these data is not shown in Fig.~\ref{fig:xmax:statistics}. The total number of events that the HiRes collaboration has used for the \Xmax analysis above \energy{18.2} is 815. However, after the application of the energy normalization (normalized to the TA energy scale) across experiments, 798 HiRes events remained with energies above  \energy{18.2} (17 events ended up with energies below this). The HiRes collaboration applies a cut on Rp to reduce the detector bias effect. This table shows the number of events before applying this Rp cut.}
  \label{tab:statistics}
\end{table}

The HiRes collaboration chooses a fluctuation estimator that differs
from the one published by Auger and Yakutsk.  Whereas the latter use
simply the standard deviation (denoted by \sigmaXmax), HiRes uses the
width  of an unbinned likelihood fit with a Gaussian to the
distribution truncated at 2 $\times$ RMS, denoted by \sigmaHiRes.

Fig.~\ref{fig:hires:xmax} shows the \meanXmaxBias and \sigmaHiRes as
measured by HiRes. The lines are the corresponding \meanXmaxBias and
\sigmaHiRes expectations for proton and iron compositions. The
different line types correspond to different models (QGSJet-01,
QGSJet-II, SIBYLL2.1).

Fig.~\ref{fig:ta:xmax} shows the corresponding \meanXmaxBias
observation and expectation for the TA experiment. Currently the TA
experiment does not have enough statistics to quantify the width of the
\Xmax distributions at the highest energies. 

Fig.~\ref{fig:xmax:statistics} shows the energy distributions and total 
number of events that survived the selection cuts at each experiment. For this Figure, the energy scales have been normalized to the TA energy scale. 
A summary of Figure~\ref{fig:xmax:statistics} is shown in Table~\ref{tab:statistics}.

The \meanXmax measurements from Yakutsk
(Fig.~\ref{fig:auger:yakutsk:xmax}), HiRes (Fig.~\ref{fig:hires:xmax})
and TA (Fig.~\ref{fig:ta:xmax}) experiments are consistent with the
QGSJet predictions for a constant proton composition at all energies
above 10$^{18}$~eV, whereas the \meanXmax measurements from the Pierre
Auger Observatory are significantly shallower than these predictions
above a few EeV (cf.\ left panel Fig.~\ref{fig:auger:yakutsk:xmax}).

At the same time, the width of the \Xmax distribution measured by
Auger gets narrower above a few EeV and the Yakutsk measurements of
the fluctuations are consistent with the Auger up to about 10$^{19}$~eV.  Yakutsk has one measurement of the \Xmax width above 10$^{19}$~eV
and it is wider than the Auger one by about 3 standard deviations
(right panel Fig.~\ref{fig:auger:yakutsk:xmax}). The \Xmax
distribution widths measured by HiRes at energies above
2~$\times$~\energy{19}, while consistent with pure proton composition have
 large statistical errors and do not definitely exclude a heavier composition (right panel Fig.~\ref{fig:hires:xmax}).

\subsection{A Toy MC to Evaluate the Performance of Different Fluctuation Measures}

The \Xmax distribution is expected to have an asymmetric long tail due
to the exponential nature of the depth of the first interaction.
Both, \sigmaXmax and \sigmaHiRes, could be under-estimated if they are
sampled with a small number of events. We have used a toy MC
simulation to evaluate the performance of both methods for the
quantification of the shower-to-shower fluctuations of \Xmax.

\begin{figure}[h]
\begin{center}
\includegraphics[width=0.31\linewidth]{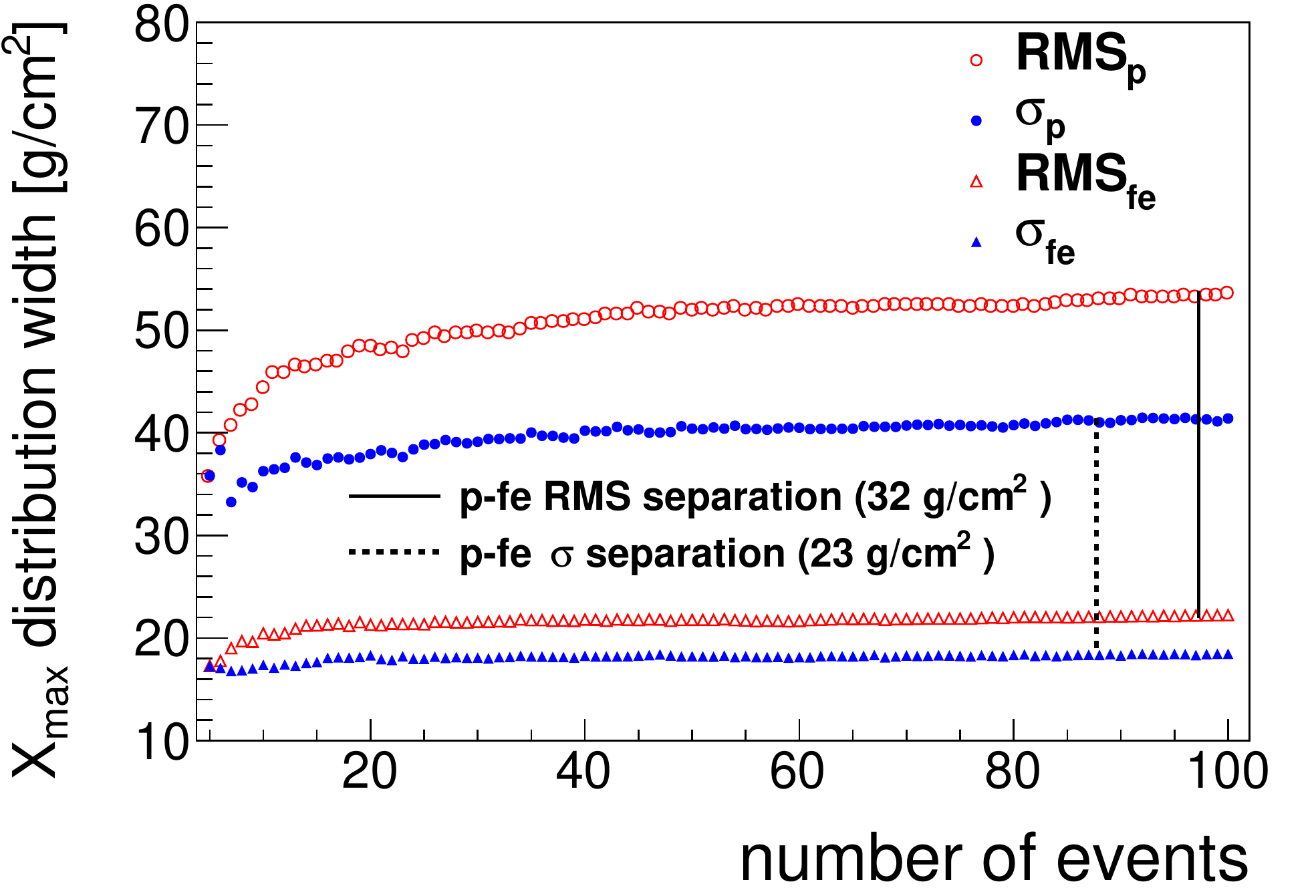}
\includegraphics[width=0.31\linewidth]{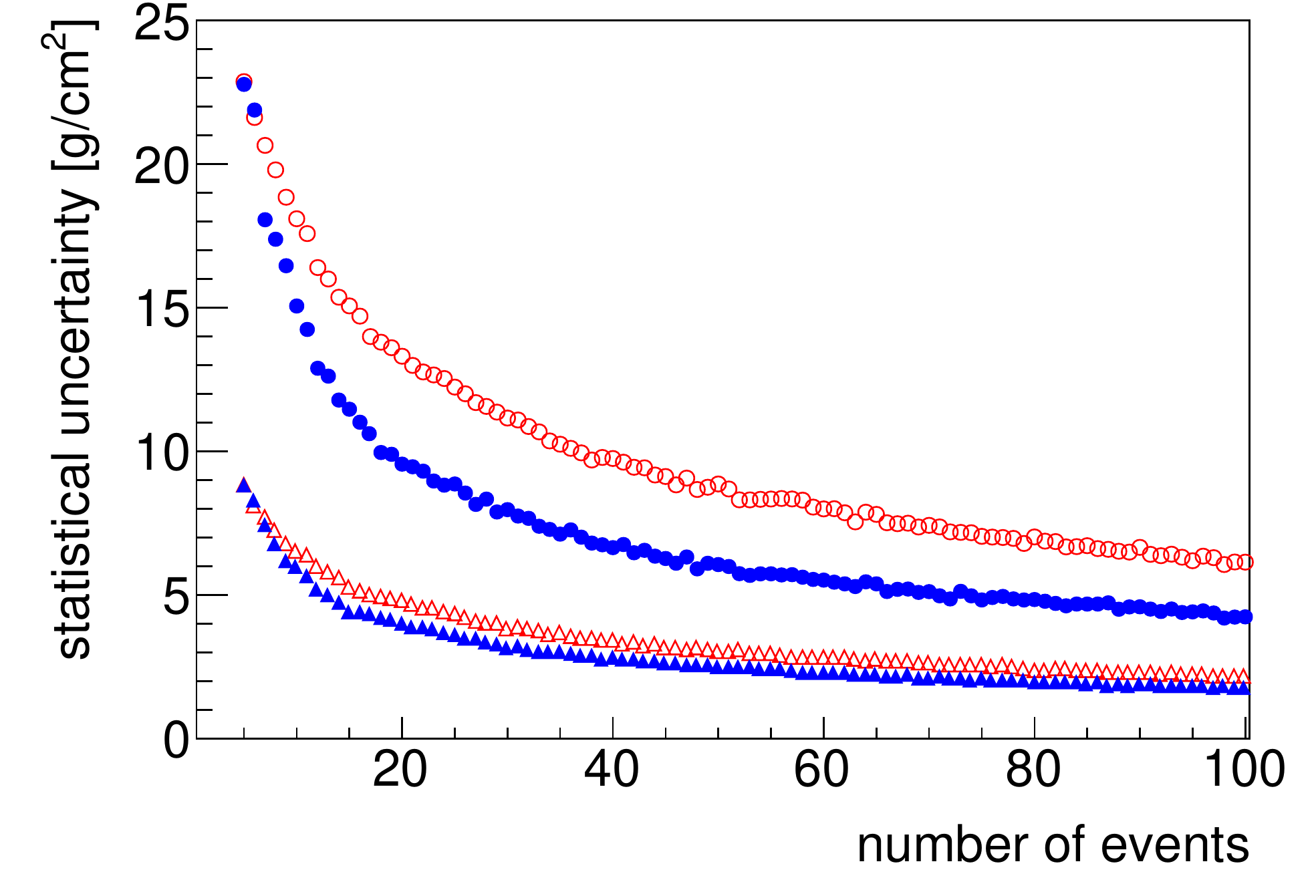}
\includegraphics[width=0.31\linewidth]{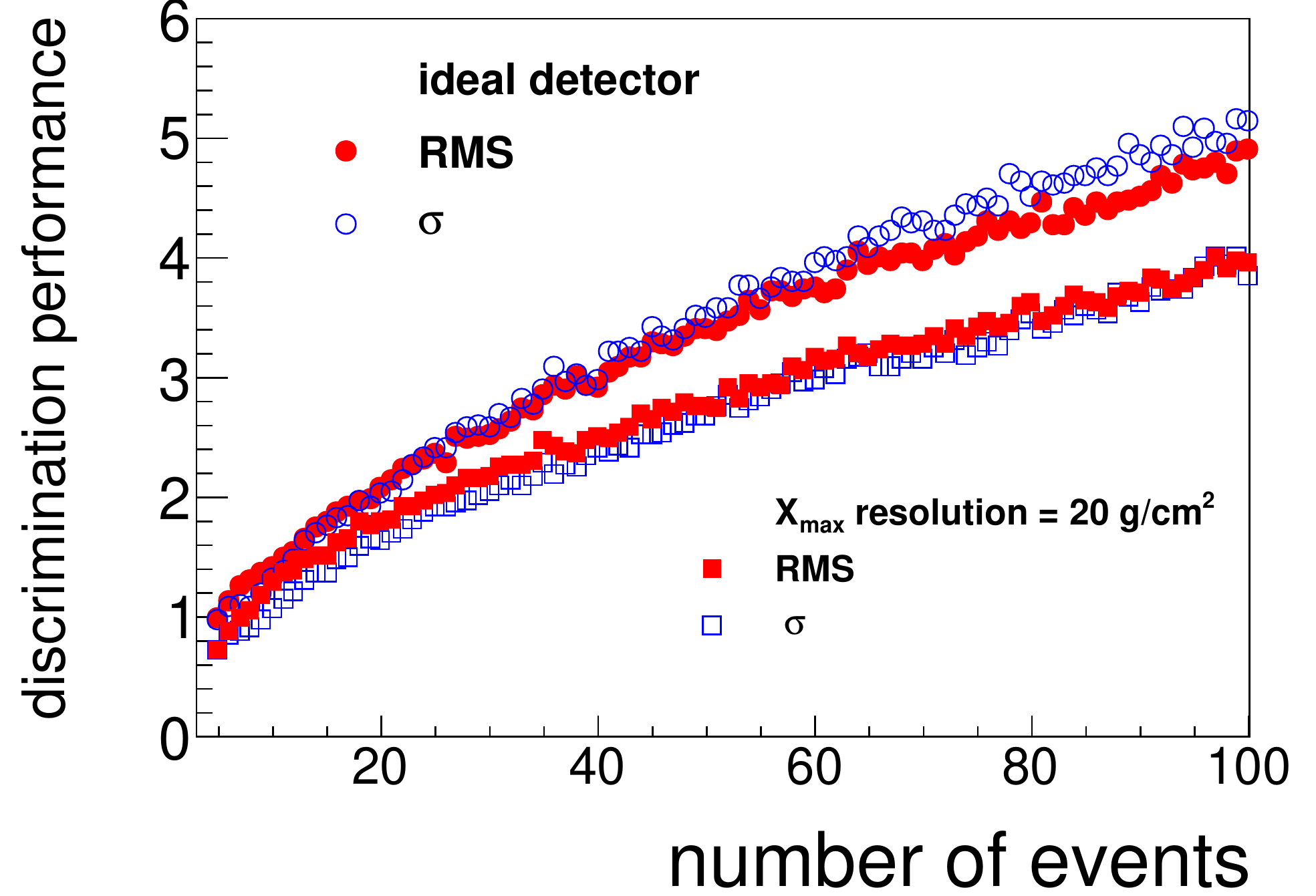}

\caption{ (left) Average value of the measured RMS and $\sigma_{X}$ as a
  function of the number of events in the sample. (middle) Statistical
uncertainty of the measured RMS and  $\sigma_{X}$. (right) Sensitivity
or discrimination power for the RMS and $\sigma_{X}$ approaches.}
\label{fig:rms:statistic}
\end{center}

\end{figure}

We have generated random samples of ``N'' events with \Xmax
distributed following the expectation for $10^{19}$~eV proton and iron
showers using the QGSJet-II model. Fig.~\ref{fig:rms:statistic} (left
panel) shows the measured RMS and $\sigma_{X}$ as a function of
the number of events ``N''. As can be seen, both approaches show a
bias in the measured \Xmax distribution width when the number of
events in the sample is small. For samples with at least 30 events,
this bias is negligible. Even for smaller statistics, the
magnitude of the bias from under-sampling is negligible compared with
the statistical uncertainty. The statistical uncertainties for the
measured RMS and $\sigma_{X}$ are shown in the middle panel of
Fig.~\ref{fig:rms:statistic}.

When using the RMS to quantify the \Xmax distribution width, the
separation between proton and iron compositions is larger. On the
other hand, the associated statistical uncertainties are smaller when
using $\sigma_{X}$. In order to evaluate the performance of using the
RMS or the $\sigma_{X}$ to quantify the width of the \Xmax
distribution, we have computed a performance indicator defined as:
\begin{equation}
\mathrm{sensitivity} =
\frac{\Delta\mathrm{width}(\mathrm{p/Fe})}{\sqrt{\sigma_\mathrm{p}^{2}
    +\sigma_\mathrm{Fe}^{2}}},
\end{equation}
where $\Delta\mathrm{width}(\mathrm{p/Fe})$ denotes the average
difference between the \Xmax distribution width for proton and iron
(measured using the \sigmaXmax or \sigmaHiRes respectively), and
$\sigma_{p}$ and  $\sigma_{fe}$ are the corresponding statistical
uncertainties of the fluctuation measurements.

Fig.~\ref{fig:rms:statistic} shows the computed
sensitivities (right panel). It turns out that the sensitivity of
both approaches is basically equivalent at all ranges of number of
events. We have also introduced a 20 g/cm$^2$ \Xmax resolution
effect to compute the sensitivity. As expected, this reduces the
sensitivity in both cases, but does not change the equivalence of
both approaches.

\section{Stability of the \Xmax Observations and Cross Checks}
\label{sec:stability}

In this section we want to show how stable the \Xmax distribution
measurements are.

\begin{figure}[h]
  \begin{center}
    \subfigure[{Auger \meanXmax}]{
      \includegraphics[width=0.31\linewidth]{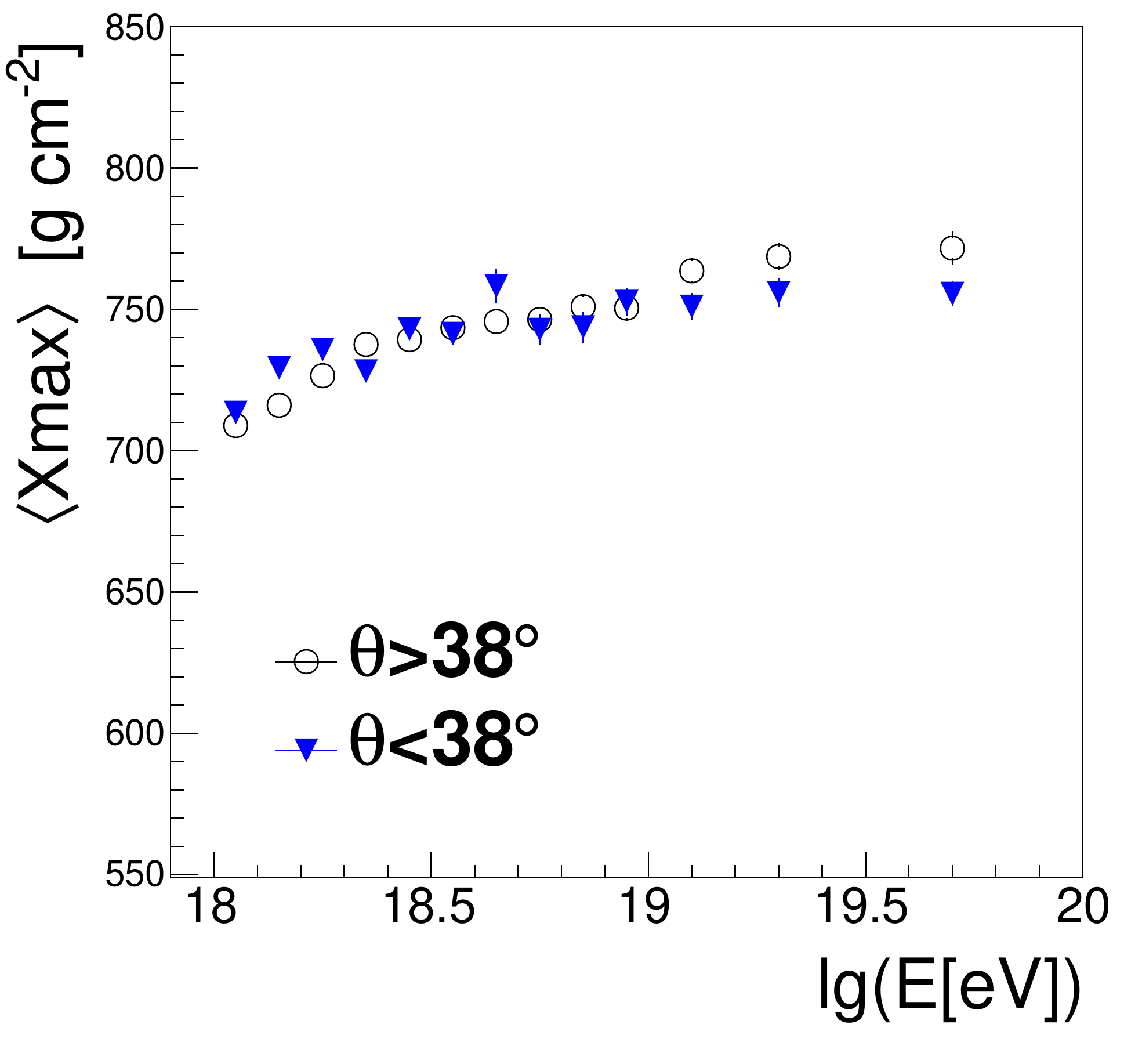}
      \label{fig:auger:xmax:zenith}
    } \subfigure[{HiRes \meanXmaxBias}]{
      \includegraphics[width=0.31\linewidth]{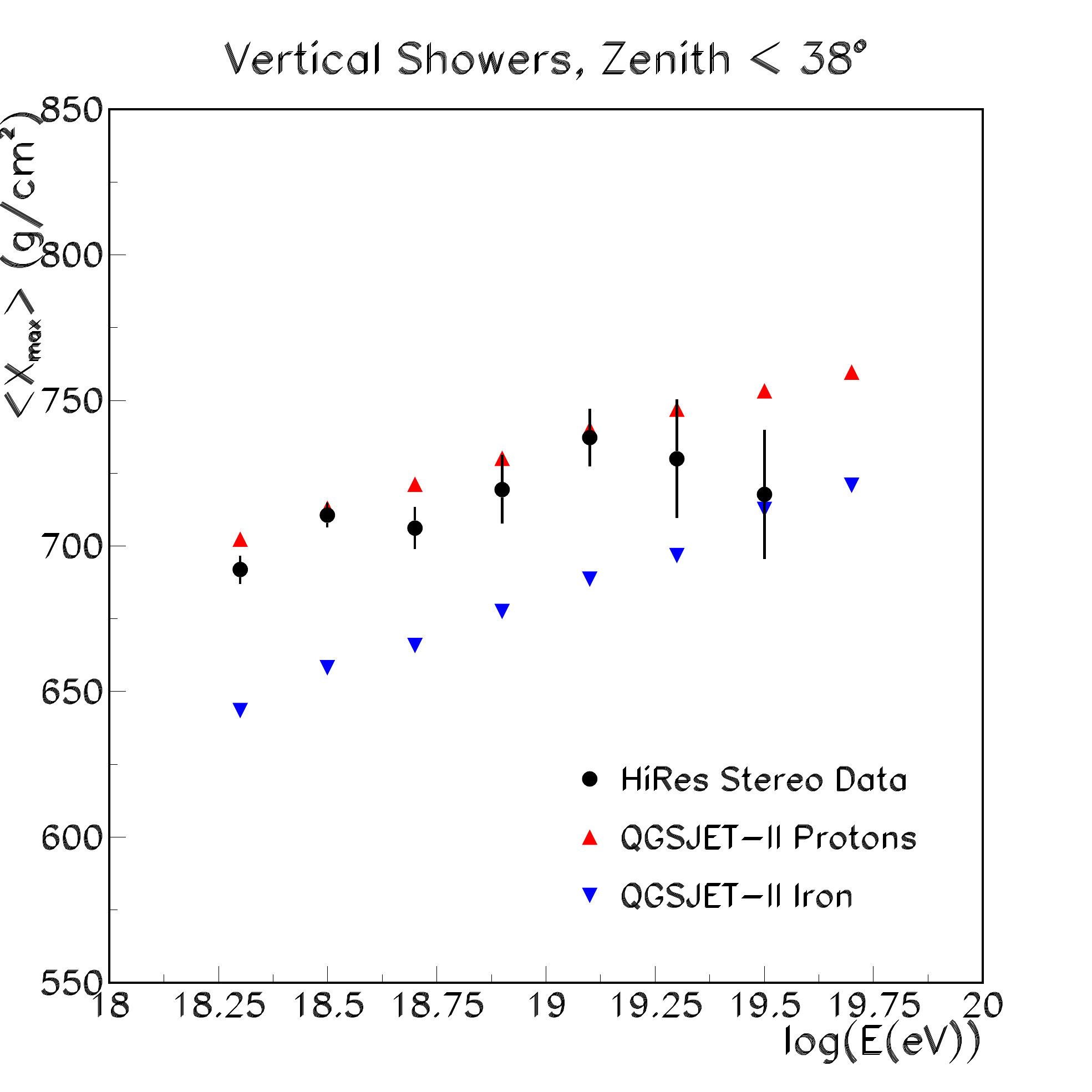}
    \label{fig:hires:xmax:vert}
    } \subfigure[{HiRes \meanXmaxBias}]{
      \includegraphics[width=0.31\linewidth]{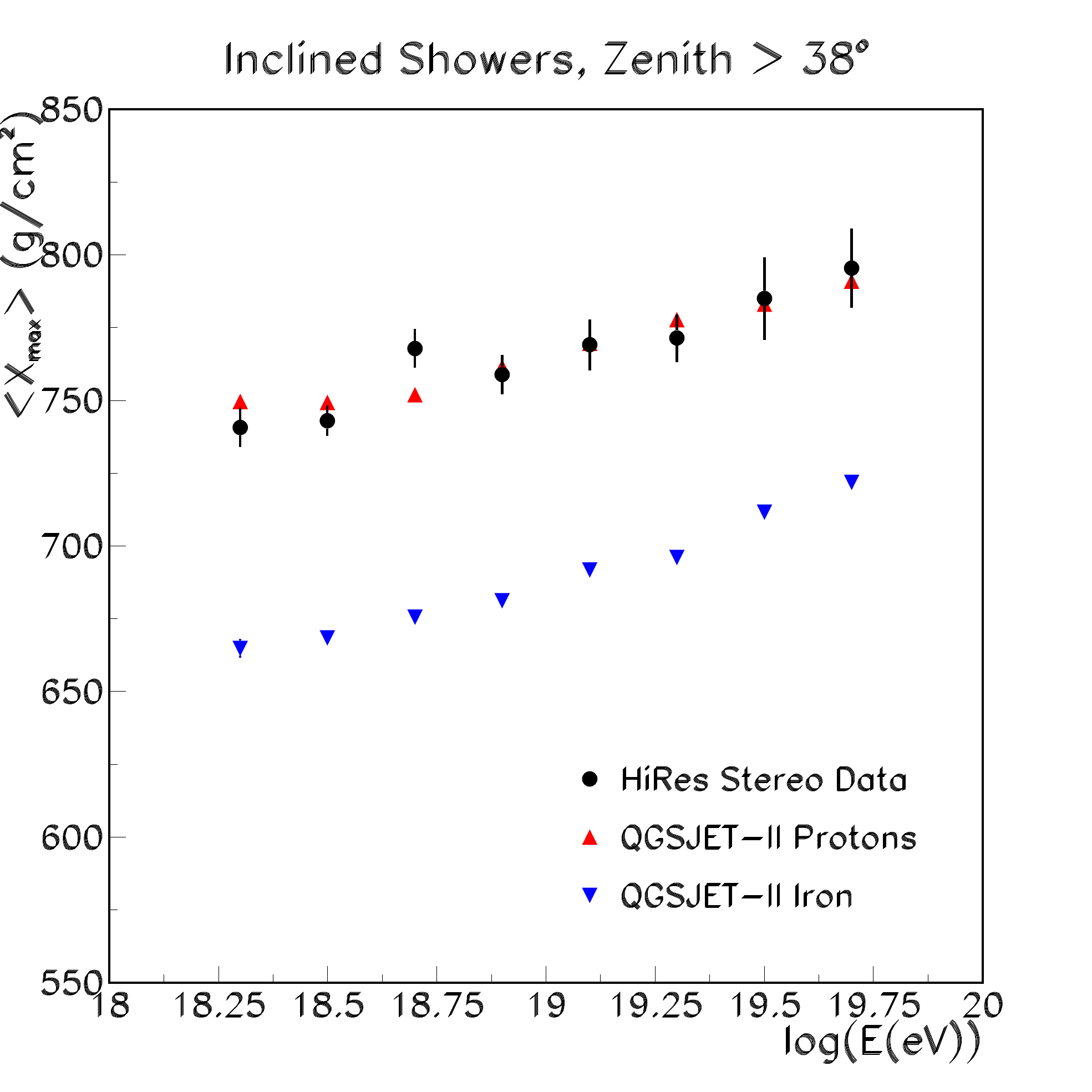}
    \label{fig:hires:xmax:incl}
    }
    \caption{The \meanXmax and \meanXmaxBias for Auger and HiRes using
      showers from different zenith angle ranges.}

    \label{fig:xmax:vert:incl}
  \end{center}
\end{figure}

We have checked whether the measured \Xmax distributions depend on the
the zenith angle. Vertical showers are more affected by the ground
level truncation of the distribution and, moreover, the fluorescence
light has to traverse denser regions of the atmosphere to the
detector.

Due to the analysis strategy used in Auger, there is no significant difference
between the vertical and inclined \meanXmax measurements, as is
illustrated in Fig.~\ref{fig:auger:xmax:zenith}).

In the case of HiRes, there is about 40~\gcm difference between the
\meanXmax measured in two zenith angle intervals, as can be seen by
comparing Figures \ref{fig:hires:xmax:vert} and
\ref{fig:hires:xmax:incl}). This difference is however well reproduced
by the detector simulation and for both zenith angle intervals the
data are compatible with the proton prediction from QGSJet-II.

 The Auger collaboration has used MC data to evaluate the  flatness of the 
detector acceptance as a function of the depth of \Xmax. Reference~\cite{bib:Vitor} shows that this acceptance becomes flat after the application of the field-of-view cuts.

\begin{figure}[h]
\centering 
{\includegraphics[width=0.48\linewidth]{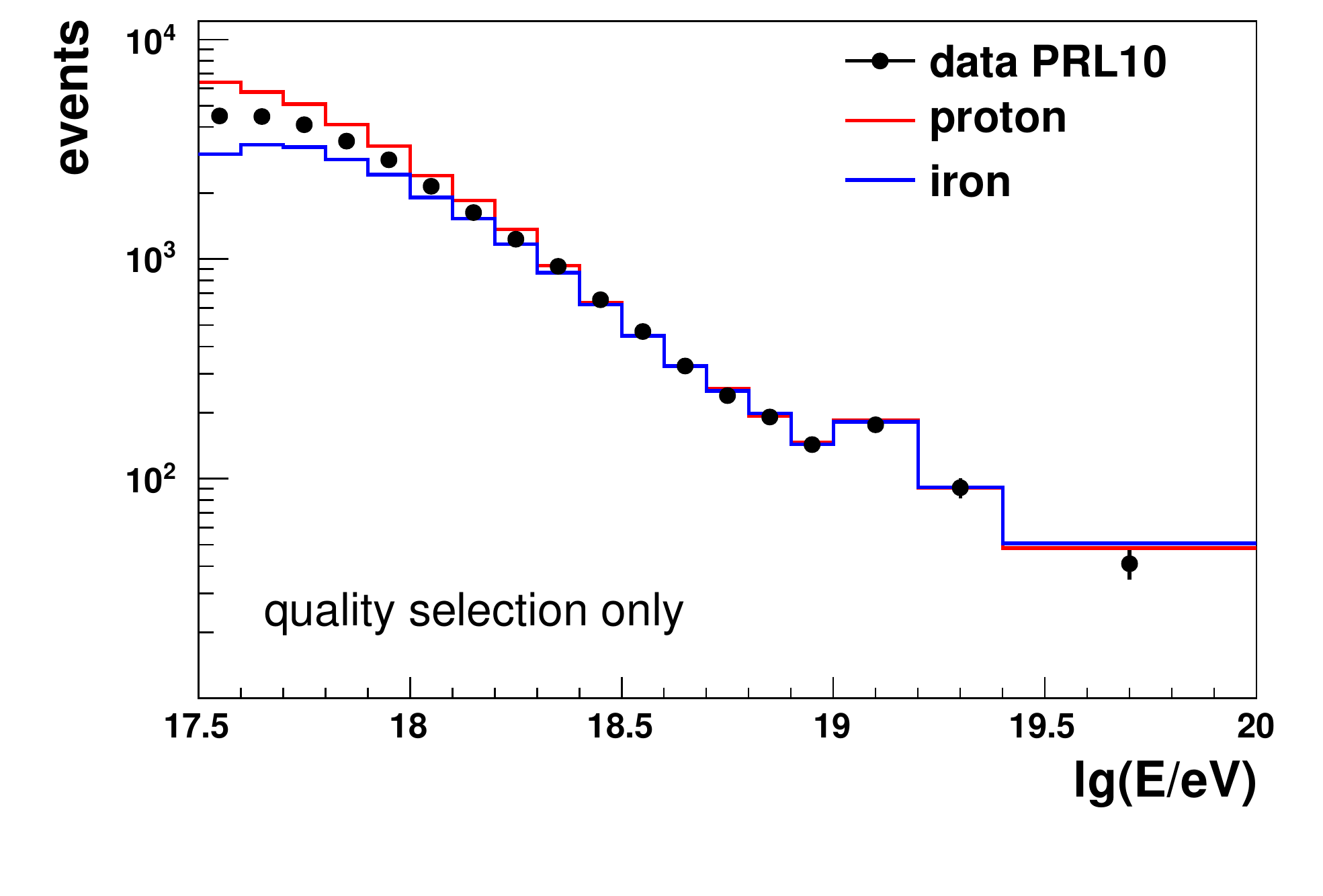}}
{\includegraphics[width=0.48\linewidth]{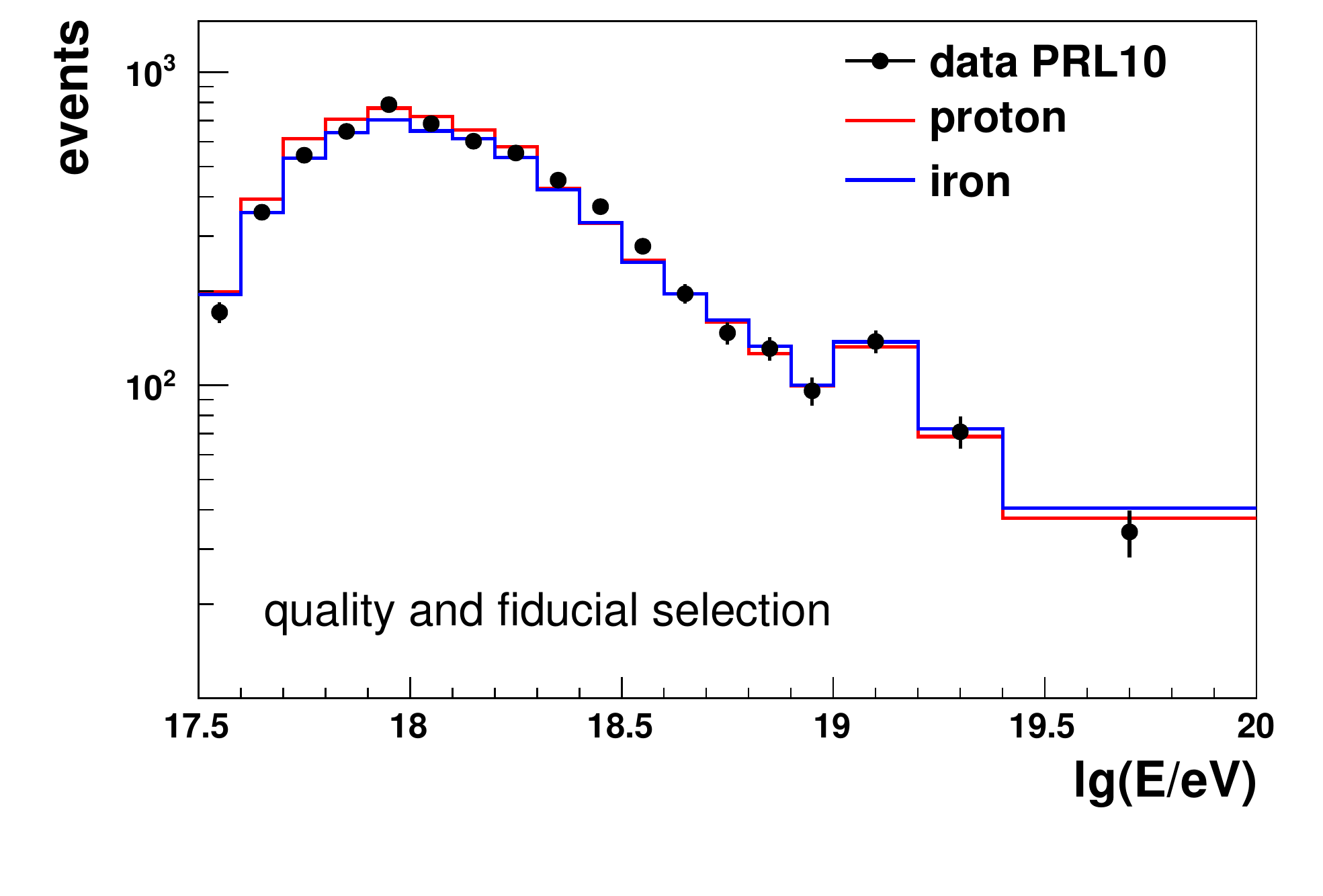}}\\
\vspace*{-1.8cm}
\subfigure[without fiducial cuts (quality cuts only)]{
\label{fig:energyDistributionQsel}
\includegraphics[width=0.48\linewidth]{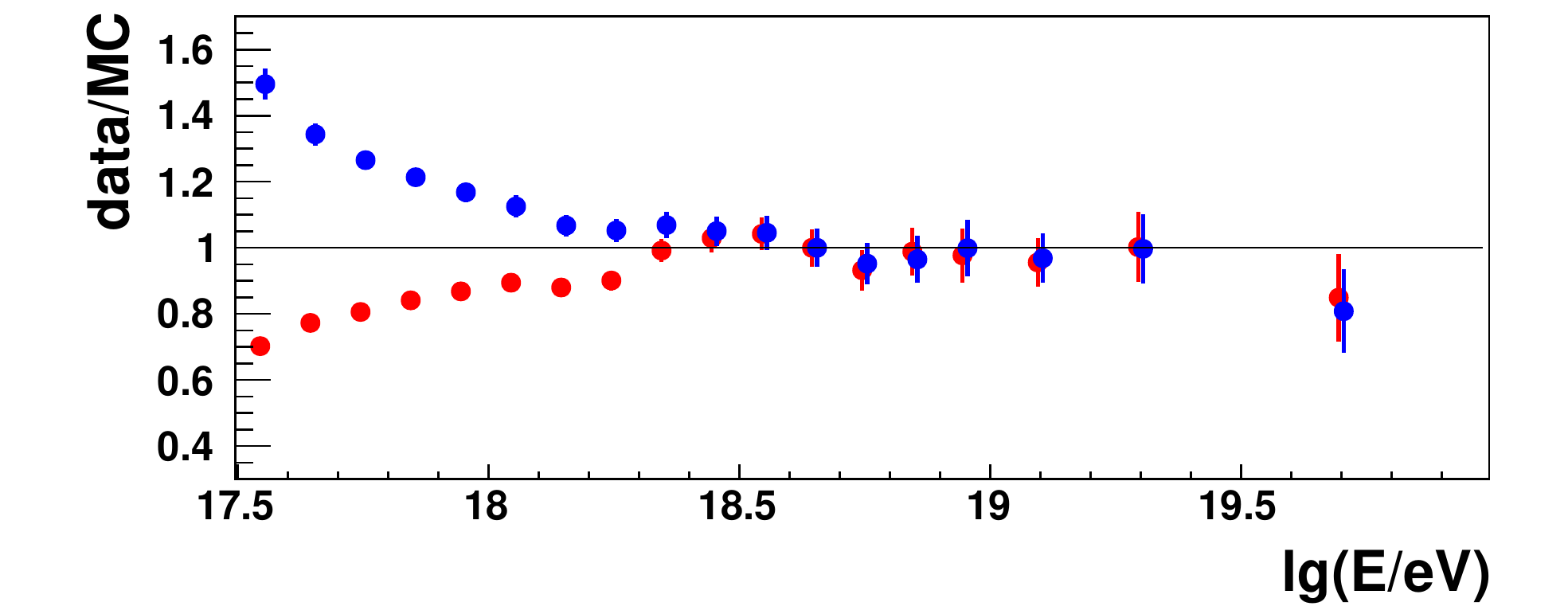}}
\subfigure[with fiducial cuts]{
\label{fig:energyDistributionQFsel}
\includegraphics[width=0.48\linewidth]{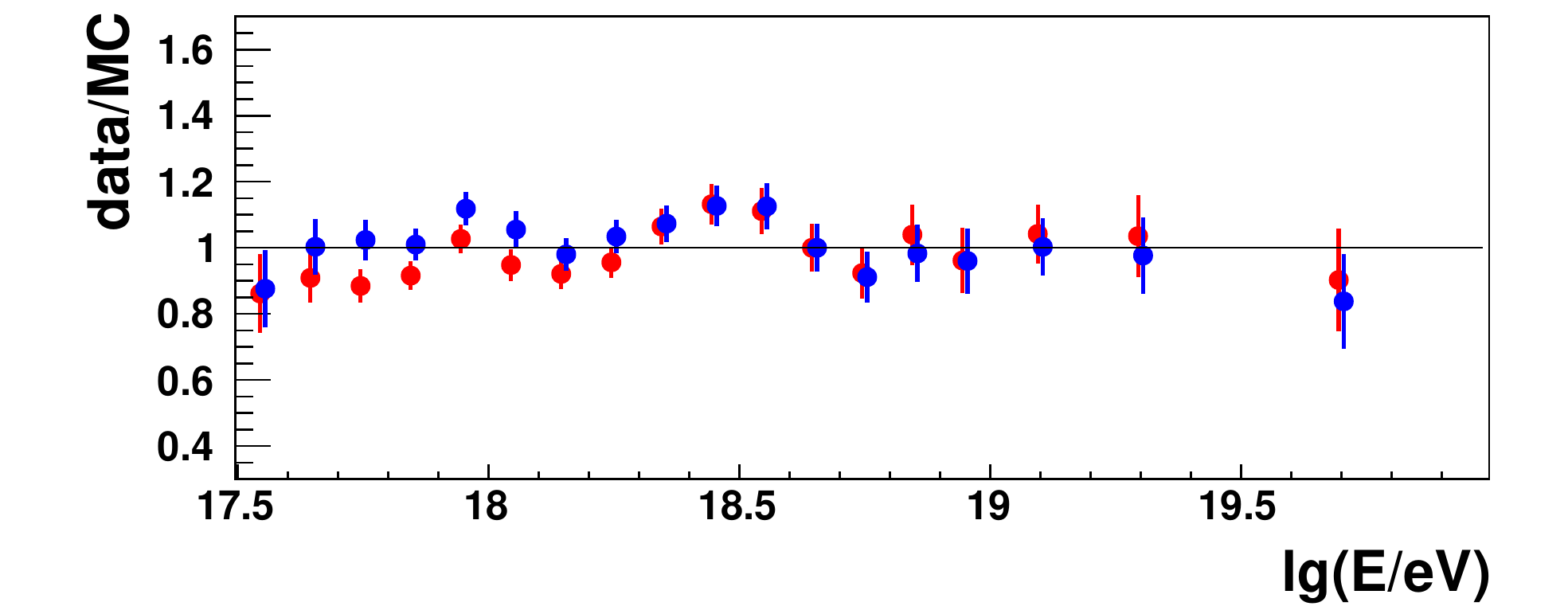}}
\caption{Energy distribution of selected events before and after
  applying the fiducial cuts. MC histograms have been normalized to
  the data at \energy{18.6}.}
\label{fig:energyDistribution}
\end{figure}

Another way to cross check that there is a homogeneous acceptance in
the Auger \meanXmax analysis, which is independent of the shower
composition, is by comparing data and MC energy distributions
resulting after the fiducial cuts. Fig.~\ref{fig:energyDistribution}
shows the energy distribution of the Auger data
from~\cite{Abraham} compared with the distributions for proton
and iron. For Fig.~\ref{fig:energyDistributionQsel}, the MC
distributions were obtained applying quality cuts only (i.e. without
applying fiducial cuts). The MC events were re-weighted at generator
level to match the spectral shape of the CR flux measured by the Auger
surface detector~\cite{Abraham:2010mj} and normalized to the data at
\energy{18.6}. As expected, the spectral shape of the MC without fiducial
selection does not match the data. It is because the acceptance
depends on the composition (or, more precisely, on the distribution of
shower maxima in the atmosphere). After application of the fiducial
field-of-view cuts, the spectral shapes of both, the proton and iron
simulations, agree well with the data
(Fig.~\ref{fig:energyDistributionQFsel}).

To check if there could be a difference in interpreting the data due
to different analysis strategies, currently the TA and Auger collaborations are 
separately working on their data interpretation using both
analysis strategies.

\section{ Validity of the Detector Monte Carlo}

The \Xmax analysis approaches followed by Auger, TA and HiRes require
some information from detector Monte Carlo simulations.

 \subsection{Auger}

For the Auger approach, the detector MC simulations are used to
estimate the average \Xmax reconstruction bias and the average \Xmax
resolution as a function of energy. They are used to correct the
observed \meanXmax and RMS(\Xmax) values respectively. After applying
fiducial volume cuts, the correction on \meanXmax is smaller than 4~\gcm, 
and the average \Xmax resolution is about 20-25~\gcm.

 The Auger collaboration has used stereo events to cross check the
 validity of its detector simulations. Stereo events have been
 simulated, reconstructed and selected in the same way as data. The
 advantage of using stereo events is that showers are reconstructed almost
 independently using each of the FD observations (they are not
 completely independent because they use the same surface station for
 the hybrid reconstruction of the geometry). As a
 result we obtain two measurements of the shower
 parameters. From the comparison of these two sets of shower
 parameters the corresponding resolutions are estimated. The
 resolution in \Xmax depends on the characteristics of the showers
 (such as geometry and energy). So, the resolution obtained using
 stereo events is not a representative \Xmax resolution of regular
 hybrid events (that are on average of lower energy than stereo
 events). However, the \Xmax resolution obtained with data and MC
 stereo events has to be consistent if the detector simulation is
 working correctly. Fig.~\ref{fig:mc:auger} shows the consistency of
 the \Xmax resolution obtained using data and simulated stereo events.

\subsection{HiRes and TA}

For the TA and HiRes approach, the detector MC simulations are used to
estimate the expected \Xmax distributions after considering the
detector effects. These expectations are estimated for different
cosmic ray primaries. Then, the expected and observed \Xmax
distributions are compared to infer the average cosmic ray
composition.

\begin{figure}[t!]
  
 \begin{center}
     \subfigure[Auger]{
       \label{fig:mc:auger}
       \includegraphics[width=0.35\textwidth]{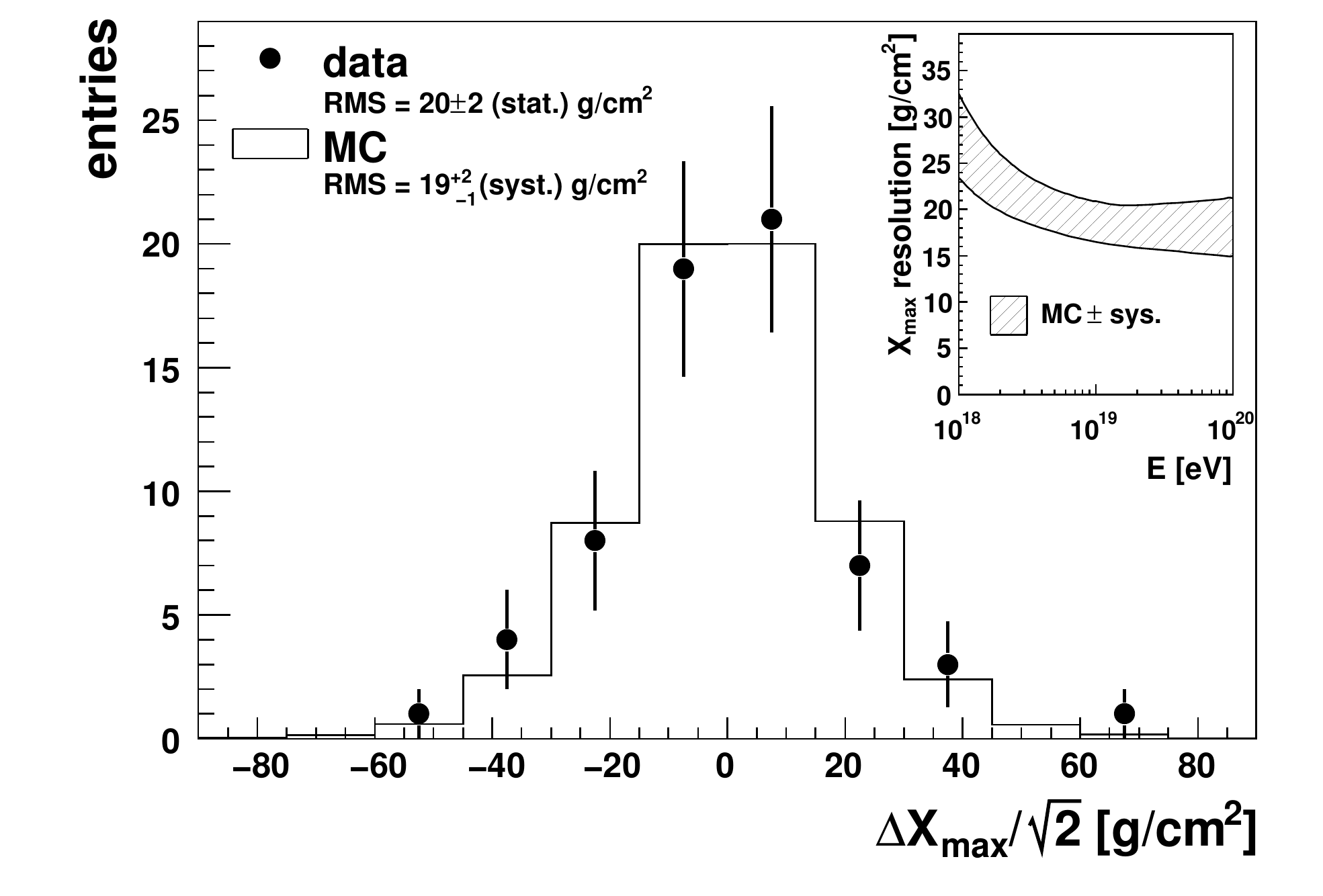}
     }
     \subfigure[TA]{
       \label{fig:mc:ta}
       \includegraphics[width=0.27\textwidth]{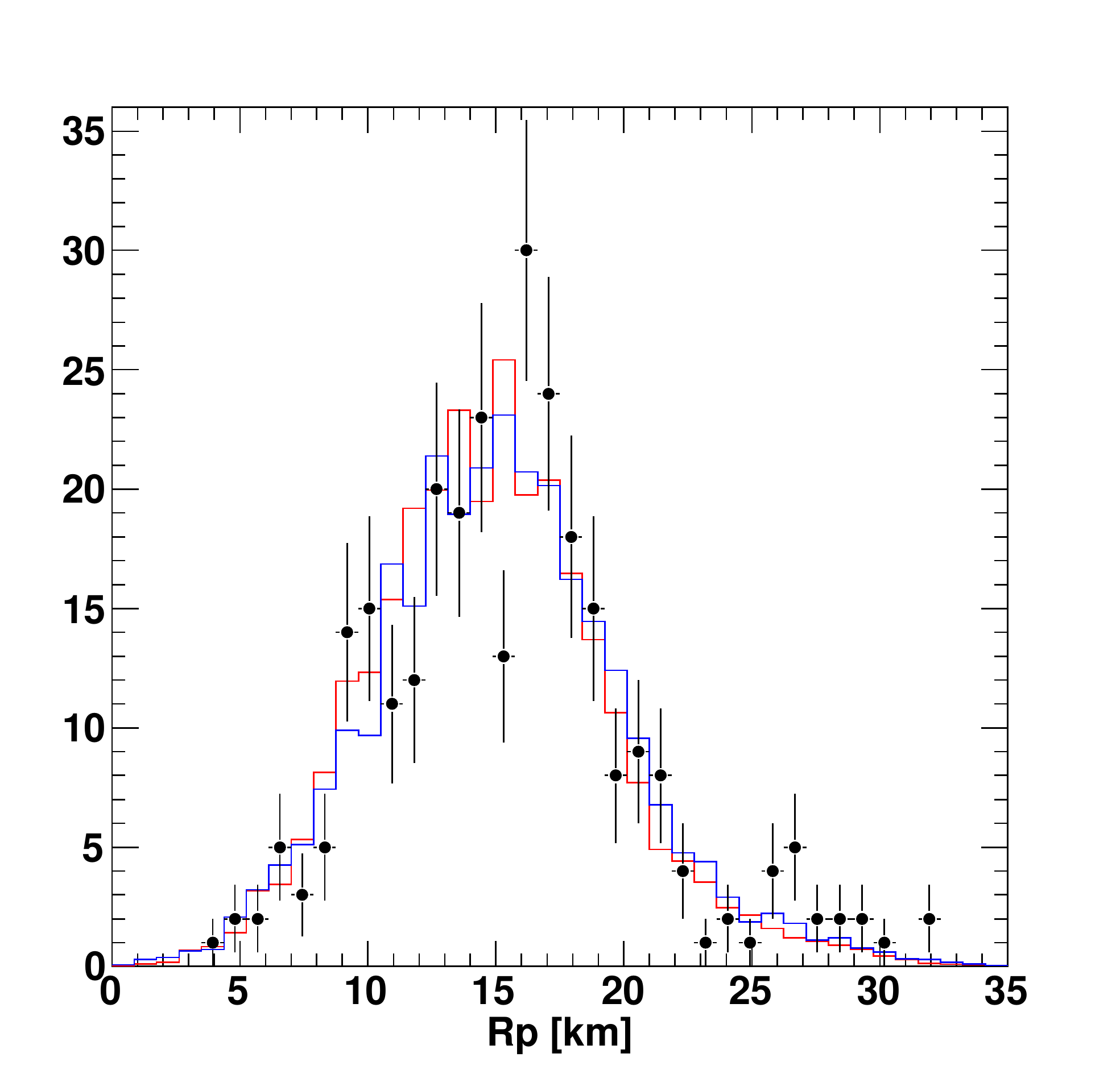}
     }
     \subfigure[HiRes]{
       \label{fig:mc:hires}
       \includegraphics[width=0.31\textwidth]{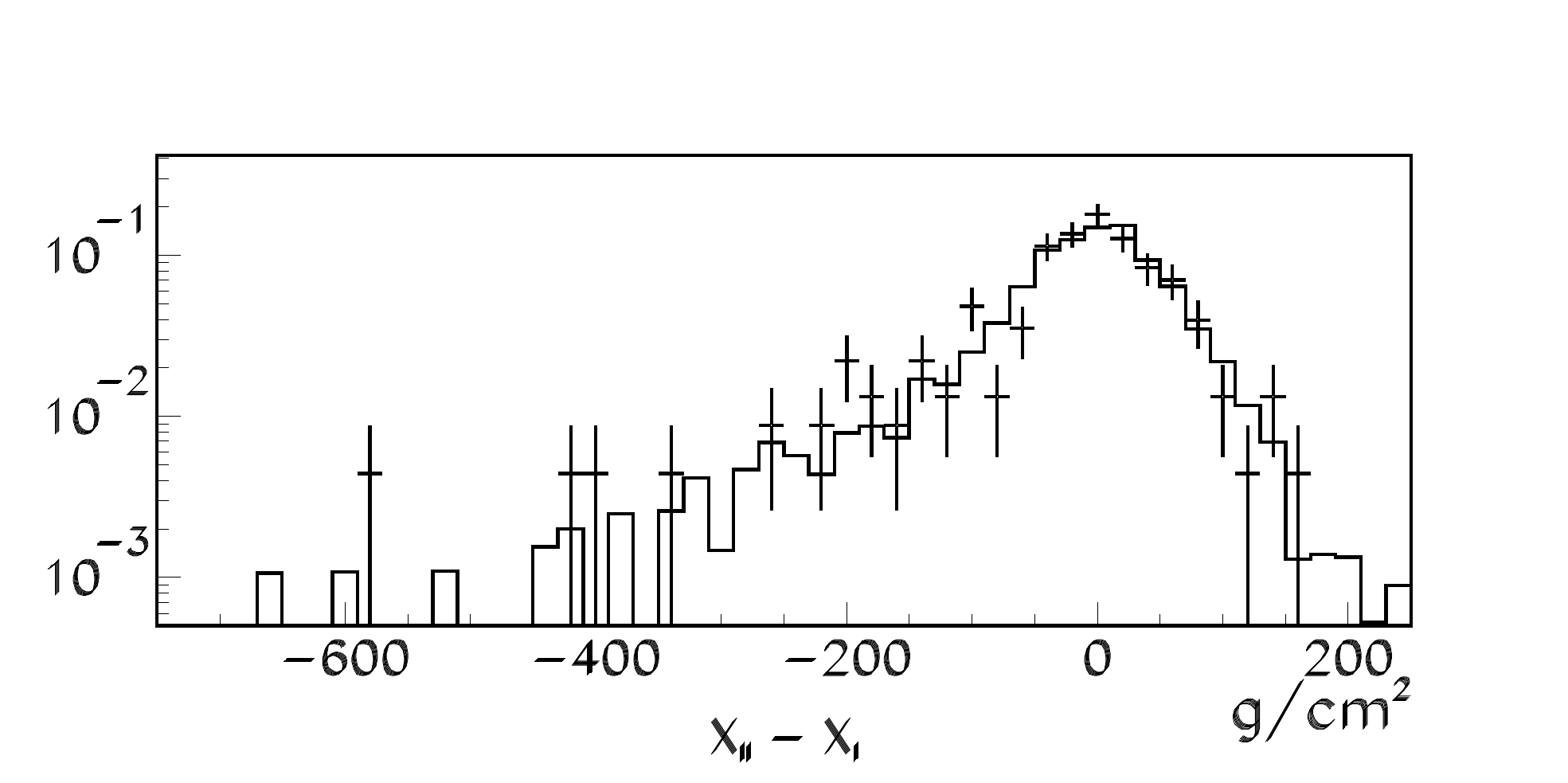}
     }

     \caption{Validating the detector MC simulations. (a) Auger \Xmax
       difference of stereo events for data and MC. The \meanXmax resolution is displayed as a function of energy in the inset.  (b) TA $R_p$ distribution for data (black points) and MC (solid lines). The red and blue lines are for QGSJet-II proton and iron composition. (c) Difference between HiRes-II ($X_{II}$) and HiRes-I ($X_{I}$) $X_{max}$ for HiRes stereo data (points) overlaid with QGSJet-II proton Monte Carlo calculations.}
     \label{fig:mc:validation}
   \end{center}
 \end{figure}

Figure~\ref{fig:mc:ta} shows the $R_{p}$ distribution for TA data and for 
MC calculations. 
Figure~\ref{fig:mc:hires} shows the \Xmax difference between HiRes-II 
($X_{II}$) and  HiRes-I ($X_{I}$) for HiRes stereo data (points) overlaid with
QGSJet-II proton Monte Carlo calculations. The asymmetry is caused by HiRes-I
covering only half the range in elevation angle. The Gaussian width of the peak
 is 44~g/cm$^2$, setting an upper limit of 31~g/cm$^2$ for the HiRes-II $X_{max}$
 resolution. Monte Carlo studies indicate that the actual HiRes-II $X_{max}$
resolution is better than 25~g/cm$^2$ over most of the HiRes energy range.

\section{ Comparison of \Xmax Results from Different Experiments}
\label{sec:xmaxComparisons}

In order to make sensible comparisons between experiments, we have
used the observed \meanXmax values to infer the average logarithmic mass, \meanlnA, using.  

 \begin{equation}
   \label{eq:lnA}
   \langle \ln A \rangle = \frac{ \langle X_\mathrm{max}
     \rangle_\mathrm{p} - \langle X_\mathrm{max} \rangle_\mathrm{data}
   } {\langle X_\mathrm{max} \rangle_\mathrm{p} -\langle
     X_\mathrm{max} \rangle_\mathrm{Fe}} \; \; \ln 56,
 \end{equation}
Similarly, one can transform \meanXmaxBias into \meanlnA by replacing
\meanXmax with \meanXmaxBias in this equation although this is only
correct as a first order approximation. This is 
because \meanXmaxBias does not correlate linearly with \meanlnA as 
\meanXmax does~\cite{bib:lorenzo}.

When we transformed the measured \meanXmax into \meanlnA, we used the
 expected \meanXmax values for proton and iron obtained directly from
 Conex simulations. On the other hand, when we transformed the
 measured \meanXmaxBias values, we used the expected \meanXmaxBias
 values for proton and iron extracted from the simulation including
 the detector.

\begin{figure}[h]
   \begin{center}
     \subfigure[using QGSJet-II model.]{
       \label{fig:lnA:qgsjetII}
       \includegraphics[width=0.48\textwidth]{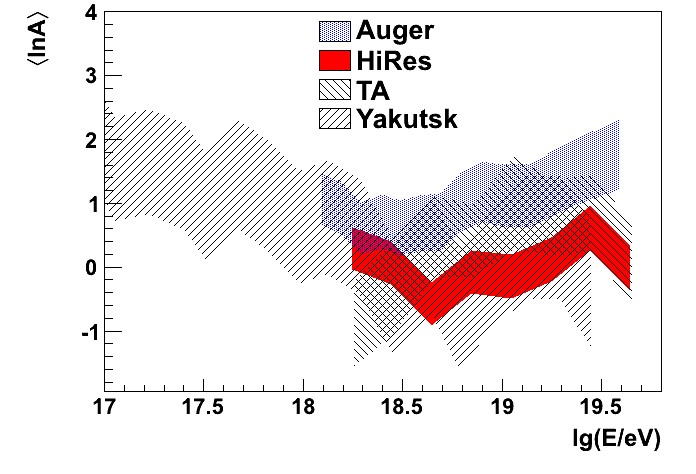}
     }
\hspace{0.0cm} \subfigure[using SIBYLL model.]{
       \label{fig:lnA:sibyll}
       \includegraphics[width=0.48\textwidth]{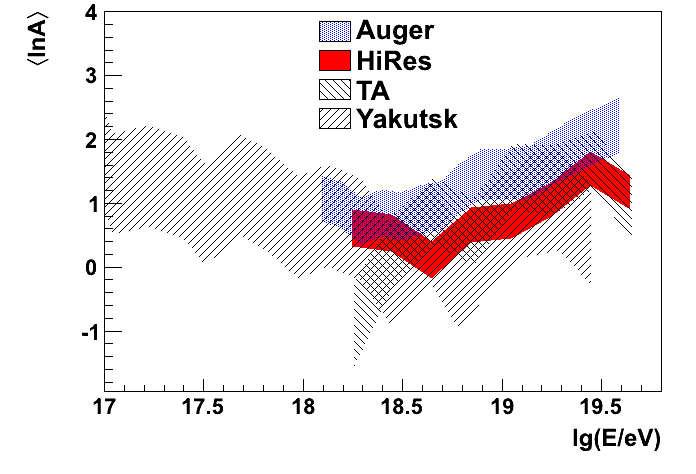}
}
     \caption{Comparing the average composition (\meanlnA) estimated
       using Auger, HiRes , TA and Yakutsk data. The shaded regions
       correspond to the systematic uncertainty ranges. To infer the average
       composition from \meanXmax, QGSJet-II and SIBYLL models have
       been used.}
     \label{fig:lnA:comparisons}
   \end{center}
 \end{figure}

Fig.~\ref{fig:lnA:comparisons} shows the \meanlnA estimated using the 
QGSJet-II and SIBYLL interaction models. The shaded regions indicate 
the range of the corresponding systematic
 uncertainties that were propagated from the systematic uncertainties
 in \meanXmax (12~\gcm  for Auger and TA, 20~\gcm for Yakutsk, and 6~\gcm 
for HiRes). 

%----------------------------------
 \begin{figure}[h!]
   \begin{center}
     \subfigure[fit to a horizontal line (constant composition).]{
       \label{fig:lnA:line:fit}
       \includegraphics[width=0.48\textwidth]{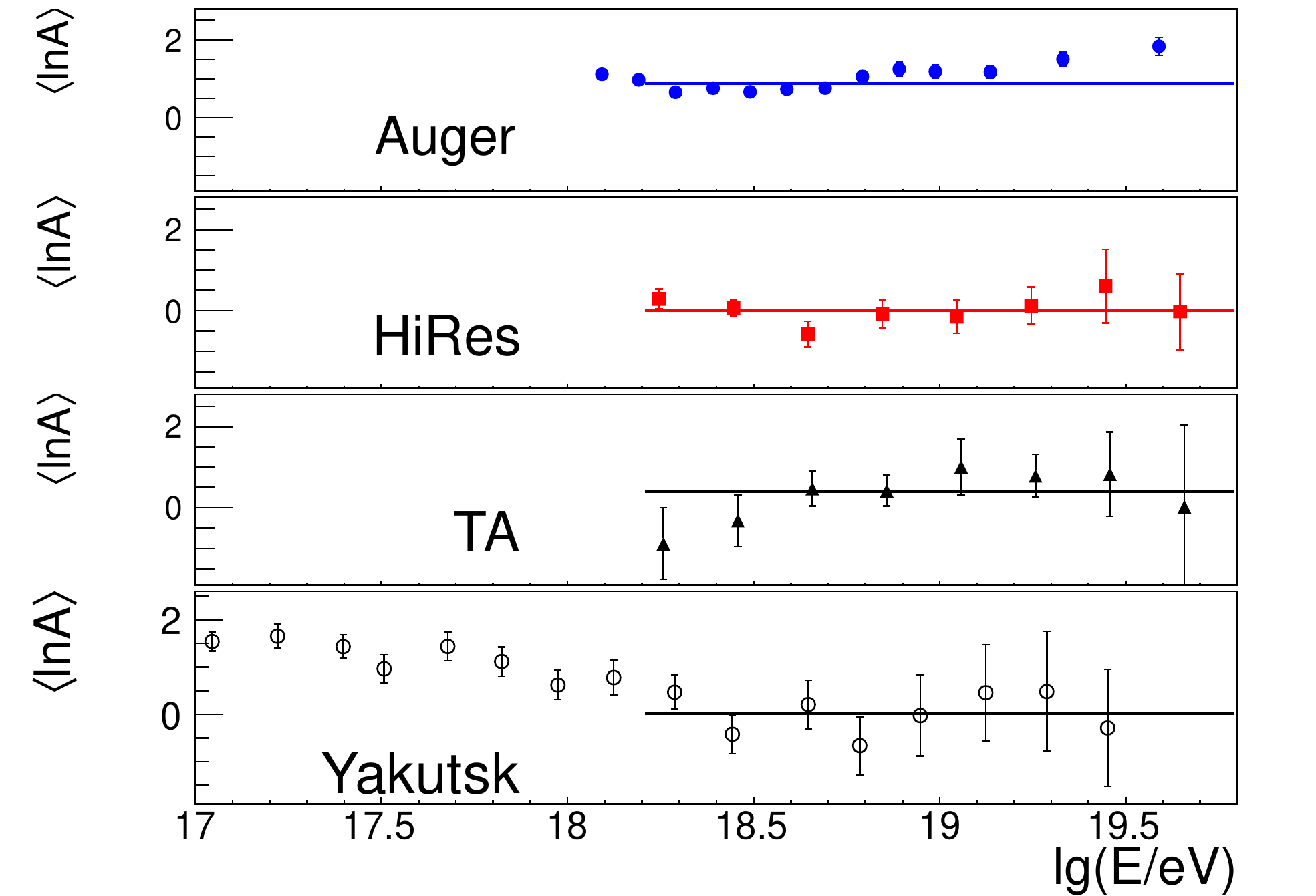}
     }
\hspace{0.0cm} \subfigure[fit to a broken line (changing
  composition).]{
       \label{fig:lnA:brokenLine:fit}
       \includegraphics[width=0.48\textwidth]{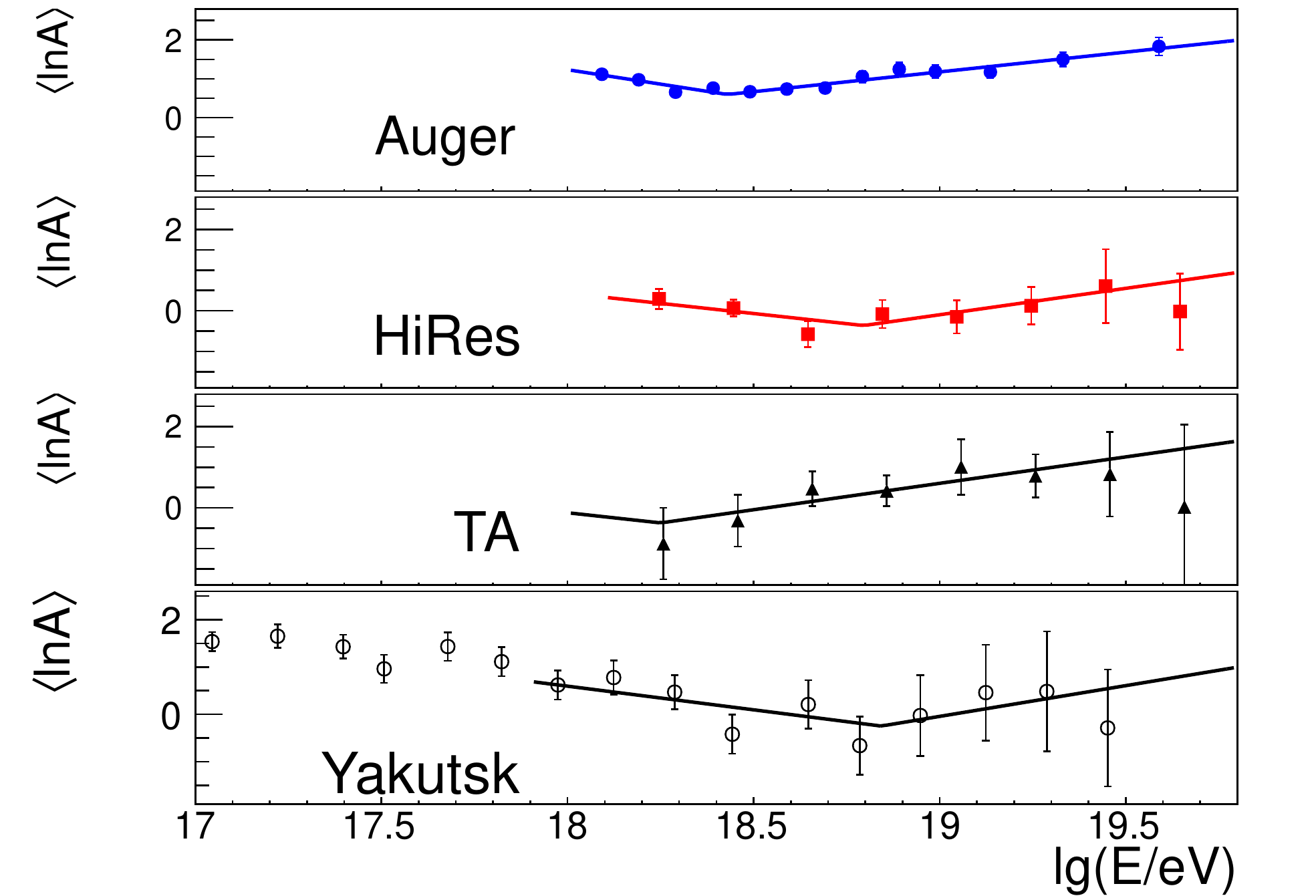}
}
     \caption{Evaluation of the average composition (\meanlnA) estimated using 
SIBYLL as a function of  energy. Two composition models are evaluated, a 
constant composition (as suggested by HiRes and TA) and a changing 
composition with a break (as suggested by Auger). The results of the fits 
are summarized in Tables~\ref{tab:lna:fit:cte1} and~\ref{tab:lna:fit:cte2}.}
     \label{fig:lnA:fits}
   \end{center}
 \end{figure}

%\vspace{-1.0cm}

\begin{table}[h!]
  \centering
  \begin{tabular}{|c|c|c|c|c|} \hline
             &  Auger & HiRes & TA  & Yakutsk \\ \hline
Constant \meanlnA   &  1.11 $\pm$ 0.03  & 0.6 $\pm$ 0.1 & 0.8 $\pm$ 0.2 & 0.3  $\pm$ 0.2 \\ \hline
$\chi^2/ndf$ & 133.6/10 & 4.4/7 & 9.8/7 & 7.7/7 \\ \hline
%Probability  & $8.48\times 10^{-24}$ & 0.72 & 0.19 & 0.35 \\ \hline
  \end{tabular}
  \caption{Fitting a horizontal line to the \meanlnA as a function of energy.}
  \label{tab:lna:fit:cte1}
\end{table}

%\vspace{-1.0cm}

\begin{table}[h!]
  \centering
  \begin{tabular}{|c|c|c|c|c|} \hline
             &  Auger & HiRes & TA  & Yakutsk \\ \hline
First Slope   &  -1.0 $\pm$  0.3 & -1.0  & -1.0 & -1.0 \\ \hline
Second Slope &  1.3 $\pm$ 0.1 & 1.3 & 1.3 & 1.3 \\ \hline
$\lg(E_{break}/eV)$ & 18.43 $\pm$ 0.04 & 18.65 $\pm$ 0.07 & 18.26 +0.14/-$\infty$   & 18.62 $\pm$ 0.14\\ \hline
$\langle \ln(A) \rangle_{break}$ & 0.75 $\pm $ 0.05 & 0.26 $\pm$ 0.10 & 0.05  +0.22/-$\infty$  & 0.08 $\pm$ 0.15 \\ \hline
$\chi^2/ndf$ & 7.4/9 & 1.23/6 & 3.37/6 & 4.22/8 \\ \hline
%Probability  & 0.59 & 0.83 & 0.76  &  0.87 \\ \hline
  \end{tabular}
  \caption{Fitting a broken line to the \meanlnA as a function of energy. For HiRes, TA and Yakutsk the slopes were fixed to the Auger ones, only the position of the breaking points were fitted}
  \label{tab:lna:fit:cte2}
\end{table}

%-------------------------------------

HiRes quotes 
 systematics broken  down into a 3.4~\gcm shift in the mean and an uncertainty 
of 3.2~\gcm/decade  in the elongation rate. For the purposes of the present 
comparison, we have combined the two HiRes uncertainties into a single 
number by adding in quadrature the uncertainty in the mean and the shift due to 
a $1~\sigma$ variation in slope over 1.6 decades of energy.

All the systematic uncertainties (on the measured \meanXmax) used in this 
work correspond to each experiment's quoted value. This working group has not 
attempted to  validate those values.

At ultra-high energies, the Auger data suggest a larger \meanlnA than
  all other experiments. The Auger results are consistent
  within systematic uncertainties with TA and Yakutsk, but not fully consistent with HiRes. HiRes is compatible with the Auger data only at energies below \energy{18.5} when using QGSJet-II (Fig.~\ref{fig:lnA:qgsjetII}), and when using SIBYLL model, Auger and HiRes become compatible within a larger energy range (Fig.~\ref{fig:lnA:sibyll}).

Comparing Figs.~\ref{fig:lnA:qgsjetII} and \ref{fig:lnA:sibyll} we
 find that the level of incompatibility between Auger and HiRes data
 depends on the model used to interpret the \meanXmax
 observations. Different models predict different ranges of \Xmax
 values for proton and iron cosmic rays, and depending on how these
 predictions compare with the range of \Xmax values that could be
 inside the FOV of the detector, the \meanXmaxBias (observed
 by HiRes) could be more or less different to the intrinsic \meanXmax,
 changing the interpretation of \meanXmaxBias. The HiRes results are 
compatible in every way with the interpretation that the
composition is light, {\em i.e.} lighter than the CNO group of elements.
 The Auger \meanXmax and \sigmaXmax results do not allow this interpretation.

 Fig.~\ref{fig:all:xmax} shows that the \meanXmax observed by Auger
 and the \meanXmaxBias observed by HiRes and TA are similar. Is there any 
physical reason that the \meanXmax for Auger and the \meanXmaxBias for HiRes and TA are all similar, or is it just coincidence?.  A direct way of
 checking the Auger and HiRes/TA compatibility would be to simulate a
 hypothetical composition which had the same \Xmax distributions as
 observed by Auger. Then this composition would be propagated through
 the HiRes and TA detector simulations and the expected \meanXmaxBias
 computed. So, we could compare directly the expected and observed
 \meanXmaxBias to evaluate the compatibility of the Auger and HiRes/TA
 observations (this is work in progress).

% An intrinsic narrow \Xmax distribution ranging within the HiRes and TA detector fields of view, will be observed free of detector bias (observed \meanXmaxBias similar to the true \meanXmax). This could be an explanation for the observed agreement in Figure~\ref{fig:all:xmax}. 

 We have also evaluated how the average logarithmic mass estimated by the
 experiments evolves as a function of energy.  Currently there are two different
models suggested by the Auger and HiRes collaborations. The \meanXmax and  
\sigmaXmax observed by the Auger experiment suggest that the composition 
might be becoming lighter with energy up to \energy{18.3}, and heavier above 
this energy. On the contrary, the  \meanXmax and  \sigmaXmax observed by the 
HiRes experiment is consistent with a constant composition (light composition) 
all along the  observed energy range.  We have evaluated both, the Auger and 
HiRes composition models using Auger, HiRes, TA and Yakutsk data (only 
statistical uncertainties were considered for this evaluation). The results 
are summarized in Tables~\ref{tab:lna:fit:cte1} and~\ref{tab:lna:fit:cte2}.

 Fig.~\ref{fig:lnA:line:fit} shows the test of the HiRes model (a fit to 
a horizontal line).  A horizontal line means constant composition in this plot. 
The large $\chi^2/ndf$ 
resulting  from the fit of the  Auger data ($\chi^2/ndf$=137/10) indicates 
that Auger data does not favor a constant composition model, but all other 
experiments have $\chi^2/ndf$ values embracing the constant composition model 
(see Table~\ref{tab:lna:fit:cte1}).

 Fig.~\ref{fig:lnA:brokenLine:fit} shows the test of the Auger model (a fit to 
a broken line). The
 fitted parameters are only the energy and \meanlnA values at which the
 lines break. The slopes before and after the breaking
 point are fixed to the results of the Auger fit. The $\chi^2/ndf$ values 
for these fits are small. However, the Auger energy and \meanlnA 
for the break point is not statistically compatible with the break points 
fitted by  HiRes, TA or Yakutsk (see Table~\ref{tab:lna:fit:cte2}). Further, 
studies (exploring the effect of different interaction models) and more 
statistics in the Northern Hemisphere are required to establish the level 
of compatibility between Southern and Northern Hemispheres.

\section{Other Observations Sensitive to Mass Composition}
\label{sec:other:par}
Apart from \Xmax observations, other shower observables can also
provide information of the average composition. Yakutsk uses an array
of muon detectors~\cite{bib:yakutsk:muons} to measure muon signals at
ground level. Auger uses its ground array of water Cherenkov tanks to
measure the signal asymmetries around the shower core, and to estimate
the muon production depth (MPD) maximum. These observations together
with the assistance of Monte Carlo simulations of the detector and
hadronic interaction models provide measurements of the average
composition (\meanlnA).

\begin{figure}[h]
   \begin{center}
     \includegraphics[width=0.6\textwidth]{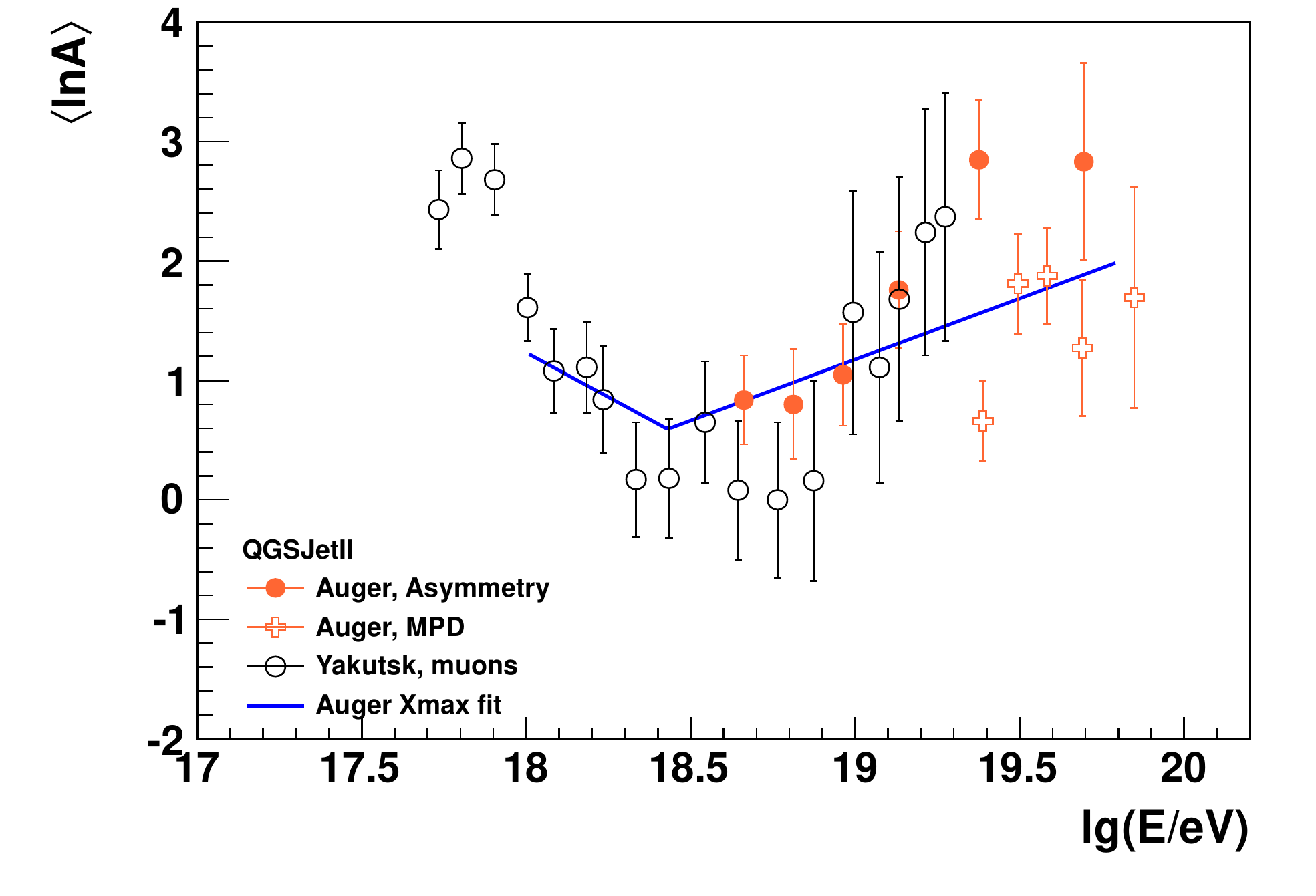}
     \caption[lnAOther]{Average composition estimated using other (other than
       \Xmax) shower observables. Open circles are using muon detectors
       from the Yakutsk experiment~\cite{bib:yakutsk:muons}, solid circles
       use the observed shower asymmetries around the core with
       the Auger SD~\cite{auger:asymmetry}, and open crosses are using
       the estimated muon production depth maximum with the Auger SD.
     }
     \label{fig:lnA:sd}
   \end{center}
 \end{figure}

Fig.~\ref{fig:lnA:sd} shows the average composition as a function of
energy estimated using the muon detectors from the Yakutsk experiment,
and the Auger ground array. For comparison purposes we have also
included the broken lines fitted to the composition estimated using the \Xmax
observations from Auger. For all these estimates
of the average composition the model QGSJet-II has been used. 

Despite some systematic difference between measurements from Auger \Xmax and Yakutsk muons, both observations suggest that the composition becomes lighter up to about \energy{18.5} and then it becomes heavier again above this energy. Measurements from Auger asymmetries also suggest that the composition becomes heavier above \energy{18.5}. Measurements from  Auger MPD only expand within a narrow energy range, and they do not provide much information regarding the evolution of the composition as a function of energy.

%\newpage

\section{Discussion}
\label{sec:discussion}

  When comparing the \meanlnA values estimated from the \meanXmax observations, results from Auger, TA and Yakutsk are compatible within systematic uncertainties. TA and Yakutsk are also compatible with HiRes. However, Auger and HiRes are not fully compatible within systematic uncertainties.  

As shown in Section~\ref{sec:xmaxComparisons}, the level of compatibility between Auger and HiRes (the two highest-statistics observatories) depends on the particular interaction model used to interpret the \meanXmax observations. Further experimental data on the high-energy hadronic interactions, {\em e.g.} from the LHC~\cite{bib:LHCResults}, would help to refine the current composition picture.

 We need more statistics in the Northern Hemisphere (about 3 times the current statistics) in order to provide a conclusive statement to whether or not the composition is changing with energy in this Hemisphere. The current data, while completely consistent with a constant light composition, cannot definitively exclude a changing composition as suggested by Auger. More statistics are also necessary to establish whether there is indeed a difference in the \sigmaXmax at higher energies between Auger and Yakutsk (Fig.~\ref{fig:auger:yakutsk:xmax}). 

 In the Northern Hemisphere HiRes has stopped data taking in 2006, however the hybrid TA observatory with a surface area of approximately 800~km$^2$ will be acquiring additional data for the next several years at least.

 Figure~\ref{fig:lnA:sd} shows \meanlnA measurements as a function of energy using different techniques. Despite the systematic differences, the measurements suggest a composition getting lighter at energies up to about \energy{18.5} and a composition getting heavier above this energy. The systematic differences between different type of measurements are very sensitive to the particular interaction model used for the interpretation. We showed the results for model QGSJet-II (in Fig.~\ref{fig:lnA:sd}), because all experiments had results available using this model.

%Author, Journal \textbf{Volume}, (year) page numbers


\begin{thebibliography}{}

\bibitem{Abraham}
  J.~Abraham {\it et al.}  [Pierre Auger Coll.],
  Phys.\ Rev.\ Lett.\  {\textbf  104}, (2010) 091101.

\bibitem{bib:lorenzo} L. Cazon and R. Ulrich. arXiv:1203.1781. 

\bibitem{bib:hires} R. Abbasi {\it et al.} [HiRes Coll.], Phys. Rev. Lett. \textbf{104}, (2010) 161101.
\bibitem{bib:ta} C. Jui {\it et al.} [TA Coll.], Proc. APS DPF Meeting
  arXiv:1110.0133.

\bibitem{bib:yakutsk:muons} L.G. Dedenko {\it et al.} J. Phys. G: Nucl. Part. Phys. \textbf{39}, (2012) 095202.


\bibitem{bib:auger} J.~Abraham {\it et al.},  [Pierre Auger Coll.], NIM A
\textbf{523}, (2004) 50.

\bibitem{bib:auger:fluor}  J.~Abraham {\it et al.},  [Pierre Auger Coll.], NIM A \textbf{620}, (2010) 227.
\bibitem{bib:HiRes:fluor} R. U. Abbasi {\it et al.}, Astropart. Phys. \textbf{23}, (2005) 157.
\bibitem{bib:ta:fluor} H. Tokuno {\it et al.}, NIM A \textbf{676}, (2012) 54-65, and NIM A \textbf{689}, (2012) 87.

\bibitem{bib:unger:xmax}M. Unger [Pierre Auger Coll.], Nucl. Phys. B, Proc. Suppl. \textbf{190}, (2009) 240.

\bibitem{bib:bellido:xmax} J. Bellido [Pierre Auger Coll.], Proc. XXth Rencontres de Blois (2009), arXiv:0901.3389.

\bibitem{bib:corsika} D. Heck and J. Knapp, Forschungszentrum Karlsruhe,
Tech. Report, (2001).
\bibitem{bib:qgs01} N. N. Kalmykov and S. S. Ostapchenko, Phys. At. Nucl.
\textbf{56}, (1993) 346.

\bibitem{bib:qgsII} S. Ostapchenko, Nucl. Phys. B, Proc. Suppl. 151, 143
(2006).

\bibitem{bib:sib} R. Fletcher et al., Phys. Rev. D \textbf{50}, (1994) 5710.

\bibitem{bib:sib2} R. Engel et al., in Proc. 26th Intl. Cosmic Ray Conference,
Salt Lake City, Utah, (1999).


\bibitem{bib:Knurenko} S.P. Knurenko et al.,
%''Longitudinal development of showers in the energy region  10^{15}-10^{17}eV.''
Proc. 27th ICRC, Hamburg, \textbf{1}, (2001), 157.
\bibitem{bib:Hillas}  A.M. Hillas, J.R. Patterson.
%'' The relation of Cerenkov time profile widths to the distance to maximum of air showers''
J.Phys.G:Nucl.Phys., \textbf{9}, (1983), 323.

\bibitem{bib:auger:xmax:icrc11} P. Facal for the Pierre Auger Coll. ICRC
  2011, arXiv:1107.4804.

\bibitem{bib:yakutsk:xmax} E.G. Berezhko et al. Astroparticle Physics, \textbf{36}, (2012) 31.

\bibitem{auger:asymmetry} D. Garcia-Pinto for the Pierre Auger Coll. ICRC
  2011, arXiv:1107.4804.


\bibitem{bib:WG:spectrum} Energy spectrum working group report, this meeting.

\bibitem{Abraham:2010mj}
  J.~Abraham {\it et al.}  [Pierre Auger Coll.],
  Phys.\ Lett.\ B \textbf{685} (2010) 239.

\bibitem{bib:Vitor} V. de Souza for the Pierre Auger Coll. this meeting.

\bibitem{bib:LHCResults} Review of modeling and description of air showers working group report, this meeting. 


\end{thebibliography}
\end{document}